\documentclass[a4paper,prx,reprint,superscriptaddress,floatfix]{revtex4-1}
\usepackage[utf8]{inputenc}
\usepackage[T1]{fontenc}
\usepackage[a4paper,centering,hmargin=1.5cm,vmargin=2cm]{geometry}
\usepackage{float}
\usepackage{graphicx}
\usepackage{amsmath}
\usepackage{mathrsfs}  
\usepackage{braket}
\usepackage{amsfonts} 
\usepackage{amssymb} 
\usepackage{csquotes}
\usepackage{booktabs}
\usepackage{xspace}
\usepackage{tikz}
\usepackage{xcolor}
\usetikzlibrary{decorations.markings}
\usetikzlibrary{positioning,calc}
\usetikzlibrary{optics}
\usepackage{siunitx}
\usepackage{scalerel}

\input{crystallographic_notations}

\usepackage{hyperref}
\usepackage{bookmark}

\definecolor{tab_blue}{HTML}{1F77B4}
\definecolor{tab_orange}{HTML}{FF7F0E}
\definecolor{tab_green}{HTML}{2CA02C}
\definecolor{tab_red}{HTML}{D62728}
\definecolor{tab_purple}{HTML}{9467BD}
\definecolor{tab_brown}{HTML}{8C564B}
\definecolor{tab_pink}{HTML}{E377C2}
\definecolor{tab_gray}{HTML}{7F7F7F}
\definecolor{tab_olive}{HTML}{BCBD22}
\definecolor{tab_cyan}{HTML}{17BECF}

\hypersetup{
  breaklinks=true,
  colorlinks=true,
  linkcolor=tab_green,
  filecolor=tab_green,
  urlcolor=tab_green,
  citecolor=tab_green,
}

\def\blsq{\scaleto{\blacksquare}{3pt}}

\newcommand{\ee}{{\rm e}}
\newcommand{\ii}{{\rm i}}
\newcommand{\dd}{{\rm d}}

\def\CC{\ensuremath{\mathbb{C}}}
\def\Id{\ensuremath{\text{Id}}}

\newcommand{\strong}[1]{\textbf{#1}}
\def\diag{\text{diag}}

\def\BZ{\ensuremath{\text{BZ}}}

\def\SUPPL{SI\xspace}

\begin{document}
\def\topfraction{1}
\def\bottomfraction{1}
\def\dbltopfraction{1}
\def\textfraction{0}

\widowpenalty10000
\clubpenalty10000

\title{Dualities and non-Abelian mechanics}
\author{Michel Fruchart}
\email{fruchart@uchicago.edu}
\affiliation{James Franck Institute and Department of Physics, University of Chicago, Chicago IL 60637, USA}
\author{Yujie Zhou}
\affiliation{Okinawa Institute of Science and Technology Graduate University, Okinawa 904-0495, Japan}
\author{Vincenzo Vitelli}
\affiliation{James Franck Institute and Department of Physics, University of Chicago, Chicago IL 60637, USA}
\date{\today}

\begin{abstract}
Dualities are mathematical mappings that reveal unexpected links between apparently unrelated systems or quantities in virtually every branch of physics~\cite{Kramers1941,Savit1980,Urade2015,Senthil2004,Louvet2015,Devetak2006,Hull1995,Maldacena1999}. Systems that are mapped onto themselves by a duality transformation are called self-dual and they often exhibit remarkable properties, as exemplified by an Ising magnet at the critical point.
In this Letter, we unveil the role of dualities in mechanics by considering a family of so-called twisted Kagome lattices \cite{Guest2003,Souslov2009,*Sun2012,Kane2013,Paulose2015,Rocklin2017,Ma2018}.
These are reconfigurable structures that can change shape thanks to a collapse mechanism~\cite{Guest2003} easily illustrated using LEGO. 
Surprisingly, pairs of distinct configurations along the mechanism exhibit the same spectrum of vibrational modes.
We show that this puzzling property arises from the existence of a duality transformation between pairs of configurations on either side of a mechanical critical point.
This critical point corresponds to a self-dual structure whose vibrational spectrum is two-fold degenerate over the entire Brillouin zone. 
The two-fold degeneracy originates from a general version of Kramers theorem that applies to classical waves in addition to quantum systems with fermionic time-reversal invariance~\cite{Kramers1930,*Klein1952}.
We show that the vibrational modes of the self-dual mechanical systems exhibit non-Abelian geometric phases \cite{Berry1984,*Wilczek1984} that affect the semi-classical propagation of wave packets~\cite{Xiao2010}.
Our results apply to linear systems beyond mechanics and illustrate how dualities can be harnessed to design metamaterials with anomalous symmetries and non-commuting responses.
\end{abstract}

\maketitle

\begin{figure}[H]
  \centering
  \includegraphics{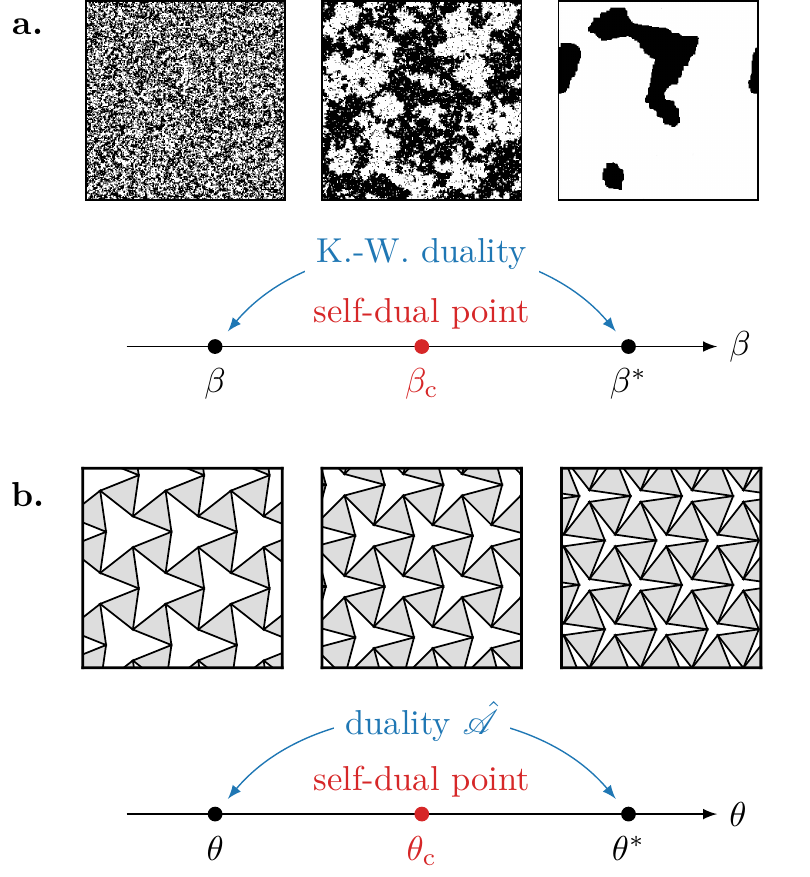}
  \caption{\label{figure_dualities_bs}\strong{Dualities.}
  (a) A celebrated example of duality due to Kramers and Wannier \cite{Kramers1941} relates the partition functions of the high-temperature and low-temperature phases of the two-dimensional classical Ising model.
  To each inverse temperature $\beta$ is associated a dual temperature $\beta^*$, and the ratio of the partition functions at $\beta$ and $\beta^*$ is a known smooth function. The self-dual point $\beta_{\text{c}} = \beta_{\text{c}}^*$ corresponds to the critical phase where the phase transition between the ferromagnet and the paramagnet occurs.
  (b) 
  Twisted Kagome lattices form a family of mechanical structures parametrized by a variable $\theta$ called the twisting angle, see Fig. \ref{figure_band_structures} for a precise definition and a LEGO model. 
  To each Kagome lattice with angle $\theta$ is associated a dual Kagome lattice with angle $\theta^* = 2 \theta_{\text{c}} - \theta$, resulting in strong relations between their mechanical properties.
  A critical point where $\theta_{\text{c}} = \theta_{\text{c}}^* \equiv \pi/4$ is observed, where the self-duality imposes strong constraints on the mechanical behavior of the mechanical structure.
  }
\end{figure}

Symmetries and their breaking are often crucial ingredients in the study and design of (meta)materials~\cite{Khanikaev2015,Susstrunk2016,Matlack2018,Bertoldi2017,Fruchart2018b,Huber2016,Cha2018}.
Dualities can be understood as a generalization of symmetries to families of theories or models~\cite{Kramers1941,Savit1980,Urade2015,Senthil2004,Louvet2015,Devetak2006,Hull1995,Maldacena1999}.
A celebrated example is the Kramers-Wannier order-disorder duality \cite{Kramers1941,Savit1980} between the low- and high-temperature phases of the two-dimensional Ising model, pictured in figure~\ref{figure_dualities_bs}a.
In this Letter, we analyze how dualities naturally emerge in the context of linear waves with a special focus on mechanics. Many mechanical structures can be effectively described as networks of masses connected by springs, even though their physical realization can be more complex~\cite{Matlack2018}. 
Their mechanical and acoustic properties are described at the linear level by their normal modes of vibration and their oscillation frequencies.
Both are determined by the dynamical matrix~$\hat{D}$ which summarizes the linearized Newton equations of motion in the harmonic approximation
$\partial_t^2 \ket{\phi} = \hat{D} \ket{\phi}$.
The vector $\ket{\phi}$ has components $\phi_p = \sqrt{m_p} u_p$, where $u_p$ is the displacement of the particle $p$ with mass $m_p$ from its equilibrium position (see SI).  
The eigenvectors $\ket{\phi_{i}}$ and eigenvalues $\omega_{i}^{2}$ of the dynamical matrix, such that $\hat{D} \ket{\phi_{i}} = \omega_{i}^{2} \ket{\phi_{i}}$, are the normal modes of vibration and the corresponding angular frequencies. In a spatially periodic system, the spectrum of the Bloch dynamical matrix $D(k)$ is organized in frequency bands with dispersion relations $\omega_i(k)$ parametrized by quasi-momenta $k$ forming the Brillouin zone of the crystal.
Although our discussion is focused on mechanics, the analysis also applies to cases where $\hat{D}$ is replaced by another linear operator, such as the Maxwell operator of a photonic crystal \cite{Joannopoulos2008}, the dynamical matrix of an electrical circuit \cite{Ningyuan2015,Albert2015,Lee2018}, or the mean-field Hamiltonian of a quantum system (in which case the eigenvalues are energies).

\begin{figure}
  \centering
  \includegraphics{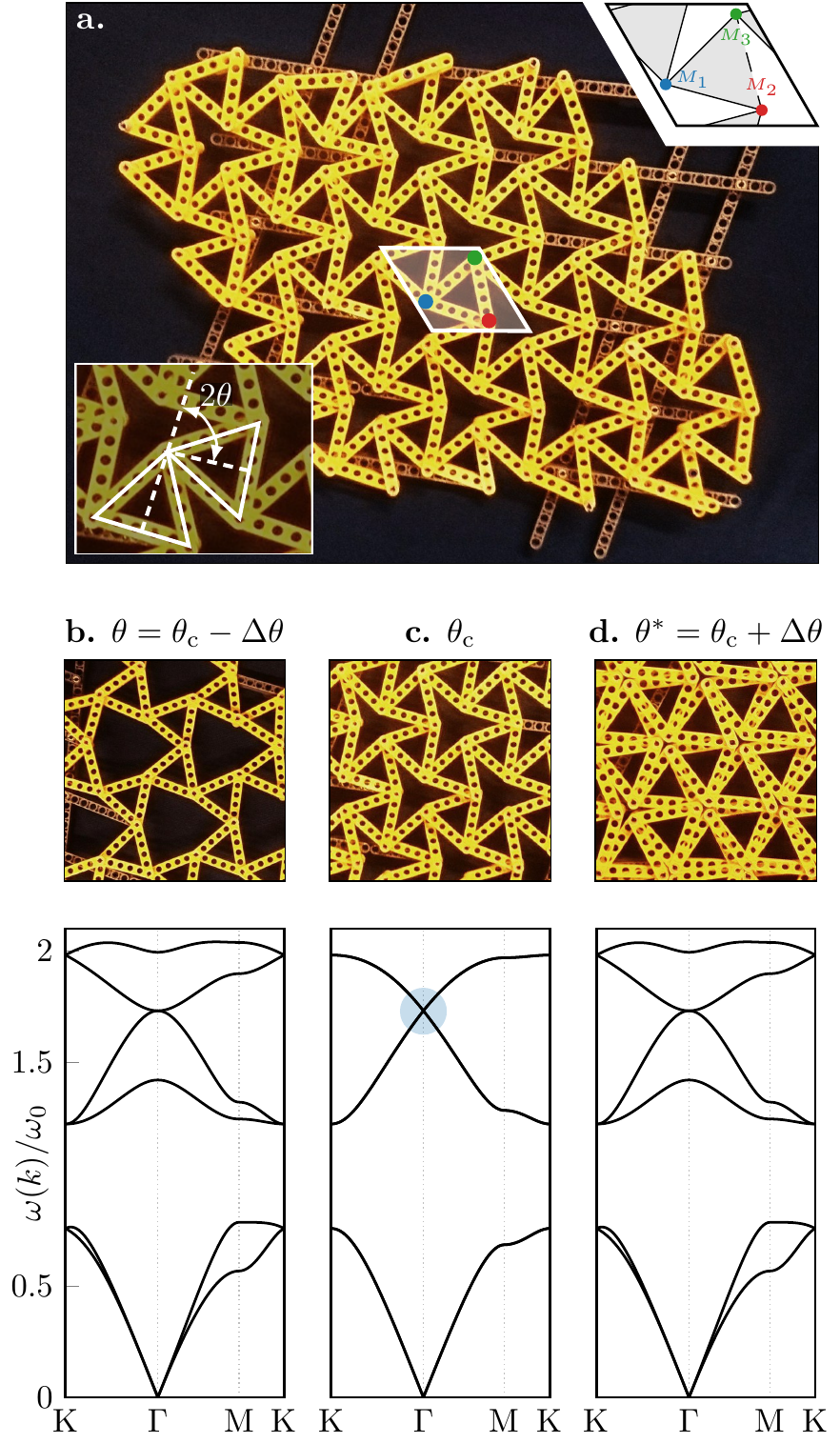}
  \caption{\label{figure_band_structures}\strong{Twisted Kagome lattices and their band structures}
  (a) A LEGO realization of the twisted Kagome lattice tuned close to the critical point $\theta_{\text{c}}$.
  Lower inset: visualization of the twisting angle $\theta$. The angle between two triangles is $\pi - 2 \theta$.
  Upper inset: unit cell of the mechanical structure. There are three inequivalent masses labeled $M_1$, $M_2$, and $M_3$.
  (b) Band structures of the mechanical structures at different twisting angles.
  The physical frequencies are nondimensionalized by a characteristic frequency $\omega_0 = \sqrt{k_0/m_0}$, 
  where $k_0$ and $m_0$ are characteristic spring constant and mass.
  The dual twisted Kagome lattices with twisting angles $\theta_{\text{c}} \pm \Delta \theta$ have the same band structure.
  The self-dual lattice with twisting angle $\theta_{\text{c}}$ has an two-fold degenerate band structure (including for points outside of high-symmetry lines).
  At the $\Gamma$ point, a double Dirac cone can be observed, highlighted by a blue disk.
  The band structures are obtained by diagonalizing the Bloch dynamical matrices $D(\theta, k)$.
  See \SUPPL for details and a movie demonstrating the collapse mechanism.
  }
\end{figure}

Twisted Kagome lattices are a family of mechanical structures obtained from a mechanical Kagome lattice \cite{Guest2003,Souslov2009,*Sun2012,Kane2013,Paulose2015,Rocklin2017,Ma2018} by actuating a mechanism, often termed a Guest-Hutchinson mode \cite{Guest2003}, that allows a global deformation of the unit cells (see SI for a movie demonstrating this property). This family is parametrized by a twisting angle $\theta$ described in Figure \ref{figure_band_structures}.
We denote by $\hat{D}(\theta)$ the dynamical matrix of the structure with the twisting angle $\theta$.
To each twisted Kagome lattice with a twisting angle $\theta$ corresponds a dual mechanical structure, which is another twisted Kagome lattice with a different dual twisting angle $\theta^* = 2 \theta_{\text{c}} - \theta$. 
Comparison of Figures ~\ref{figure_band_structures} (b) and (d) reveals that two lattices related by a duality transformation share the same band structure despite their clear structural difference. 
Remarkably, there is a self-dual Kagome structure with angle $\theta_{\text{c}}^* = \theta_{\text{c}} = \pi/4$, where the band structure is doubly degenerate, as show in Figure~\ref{figure_band_structures}c.
We now prove that the explanation of these phenomenological observations can be traced to the existence of a mathematical duality between the dynamical matrices of pairs of Kagome lattices.

\begin{figure*}
  \centering
  \includegraphics{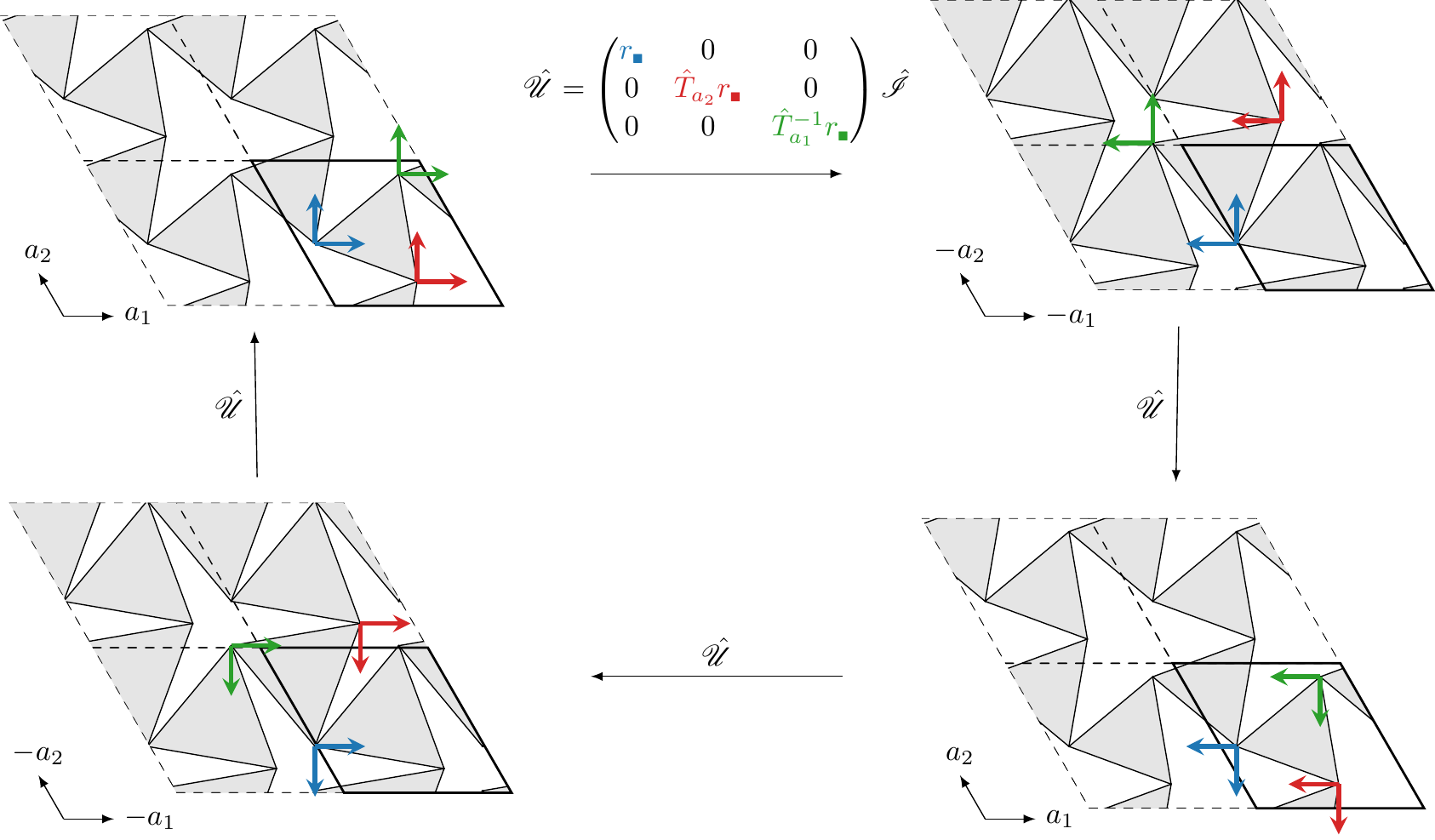}
  \caption{\label{figure_duality_operators}\strong{Schematic action of the duality operator.}
  The duality operator maps the vibrational degrees of freedom of a twisted Kagome lattice to the vibrational degrees of freedom of the dual Kagome lattice.
  The vibrational degrees of freedom (in blue, red, and green) in a unit cell (highlighted in bold) are rotated by \ang{90} counterclockwise and translated to another unit cell. Importantly, the translation depends on the degree of freedom: the vibrations of mass $M_1$ (in blue) are not shifted, while the vibrations of mass $M_2$ are shifted by one lattice vector $a_2$ and the vibrations of mass $M_3$ (in green) are shifted by another lattice vector $a_1$.
  The operator $\hat{\mathscr{U}}$ is written as a block matrix; the different blocks describes the different masses in the unit cell (as represented by the colors), and the (real) matrices $r_{\blsq} \equiv \ii \sigma_y$ acts for each mass on the two orthogonal vibrations along $x$ and $y$, mapping $(u_x, u_y)$ to $(u_y, -u_x)$. 
  The operator $\hat{\mathscr{I}}$ acts on the Bravais lattice as space inversion, but does not modify the internal degrees of freedom (see SI).
  Iterated applications of $\hat{\mathscr{U}}$ show that $\hat{\mathscr{U}}^{2} = - \Id$ and $\hat{\mathscr{U}}^{3} = - \hat{\mathscr{U}}$ while $\hat{\mathscr{U}}^{4} = \Id$, showing that the symmetry has order four.
  In the self-dual lattice, the transformation resembles a non-symmorphic symmetry composed of a \ang{90} rotation followed by a non-integer lattice translation at first sight. However, further inspection shows that this operation is different from the duality operation, and is not a symmetry of the self-dual lattice (see SI for a visual proof).
  }
\end{figure*}

A celebrated theorem from Kramers~\cite{Kramers1930,*Klein1952} states that the energy states of time-reversal invariant systems with half-integer spin are at least doubly degenerate.
At first sight, this theorem does not apply here, as the mechanical degrees of freedom are neither quantum mechanical nor fermionic.
However, Kramers theorem can still formally apply to the mechanical system, provided that we find an anti-unitary operator squaring to minus the identity that commutes with the dynamical matrix.
Here, we show how to construct such an anti-unitary operator.
To so so, we first introduce a unitary transformation $\hat{\mathscr{U}}$ acting on the vibrational degrees of freedom of a twisted Kagome lattice as represented in figure~\ref{figure_duality_operators}.
A direct calculation (see \SUPPL) shows that
\begin{equation}
  \label{unitary_nonlocal_duality}
  \mathscr{U}(k) D(\theta^*, -k) \mathscr{U}^{-1}(k) = D(\theta, k)
\end{equation}
where $\mathscr{U}(k)$ is the Bloch representation of the operator $\hat{\mathscr{U}}$.
Hence, $\hat{\mathscr{U}}$ should be viewed as a linear map between \emph{different spaces}, describing respectively the vibrations of the different mechanical structures with twisting angles $\theta^*$ and $\theta$ (compare the two lattices in figure~\ref{figure_duality_operators}).
Note that $\mathscr{U}(k)$ does not depend on the twisting angle $\theta$.
As Newton equations are real-valued, the Bloch dynamical matrix satisfies $\Theta D(\theta, k) \Theta^{-1} = D(\theta, -k)$ where $\Theta$ is complex conjugation.
Hence, by combining the anti-unitary operator $\hat{\Theta}$ with $\hat{\mathscr{U}}$, we get the desired anti-unitary operator $\mathscr{A}(k) = \mathscr{U}(k) \Theta$ which squares to $\mathscr{A}(k)^2 = - \Id$, and such that
\begin{equation}
  \label{duality_relation}
  \mathscr{A}(k) D(\theta^*,k) \mathscr{A}^{-1}(k) = D(\theta,k).
\end{equation}

Equation~\eqref{duality_relation} is the expression of a duality between the two lattices with twisting angles $\theta$ and $\theta^*$, illustrated in figure~\ref{figure_dualities_bs}b.
The dynamical matrices of the two dual systems are related by a anti-unitary transformation. As a consequence, they have identical band structures (in terms of eigenvalues; compare figure~\ref{figure_band_structures} b. and c.; more precisely, the eigenvalues are related by complex conjugation, and are equal because they are also real) and the eigenvectors are related by $\hat{\mathscr{A}}$.
Equation~\eqref{unitary_nonlocal_duality} is also a duality between the same lattices. In contrast with~\eqref{duality_relation}, it is ruled by a unitary operator, but is non-local in momentum space (it relates $k$ to $-k$). Alone, it would ensure that the band structures of both lattices are the same only up to an inversion of momentum.

At the critical twisting angle $\theta_{\text{c}} = \theta_{\text{c}}^* \equiv \pi/4$, the mechanical structure is self-dual. The duality \eqref{duality_relation} acts as a hidden symmetry of the critical dynamical matrix $D(\theta_{\text{c}})$, through $\mathscr{A}(k) D(\theta_{\text{c}},k) \mathscr{A}^{-1}(k) = D(\theta_{\text{c}},k)$.
As $\hat{\mathscr{A}}^2 = - \Id$, Kramers theorem~\cite{Kramers1930,*Klein1952} can be applied, and implies that the band structure is globally two-fold degenerate, at every point $k$ of the Brillouin zone, as observed in figure~\ref{figure_band_structures} c.
Interestingly, $\hat{\mathscr{A}}$ acts in the same way as the combination of spatial inversion and a so-called fermionic time-reversal would in an electronic system, although neither are present in our mechanical system.
Due to the presence of the self-dual symmetry, the critical band structure exhibits exotic features. To begin with, a finite-frequency linear dispersion (a double Dirac cone) is observed at the center of the Brillouin zone (called $\Gamma$ ; see figure~\ref{figure_band_structures}c.) that is uncommon in systems with time-reversal invariance~\cite{Sakoda2011,Huang2011} (see SI for a discussion). 

When self-duality is combined with the usual crystal symmetries, anomalous point groups can be realized. 
Consider paving the two-dimensional plane with a single regular polygon.
This is possible with a triangle, a square, or a hexagon, but not with a pentagon or a dodecagon.
This is a manifestation of the crystallographic restriction theorem: the only point group symmetries compatible with lattice translations are of order $1$, $2$, $3$, $4$, or $6$, in two dimensions. 
(The order of an operation~$g$ is the smallest integer~$n$ such that~$g^n$ is the identity.)
The point group \sch{C3v} of twisted Kagome lattices at the center $\Gamma$ of the Brillouin zone contains $3$-fold rotations (as visible in figure~\ref{figure_band_structures}), perfectly compatible with this assertion.
At the critical angle $\theta_{\text{c}}$, the duality relation \eqref{unitary_nonlocal_duality} turns into an additional symmetry of the dynamical matrix. 
Hence, the point group at $\Gamma$ has effectively to be supplemented with $\mathscr{U}(\Gamma)$, which has order $4$ (see figure~\ref{figure_duality_operators}). 
Combined with a $3$-fold rotation from \sch{C3v}, the self-dual symmetry $\mathscr{U}(\Gamma)$ produces an anomalous symmetry of order $3 \times 4 = 12$ making the effective point group at $\Gamma$ non-crystallographic (isomorphic to $D_{12}$, see \SUPPL).
The emergence of this non-crystallographic point group is curious, as the twisted Kagome lattices are indeed crystals, not quasicrystals. 
However, there is no contradiction with the crystallographic restriction theorem, because the self-dual symmetry is not a spatial symmetry.

We now show how to generate non-Abelian sound in our self-dual mechanical structures. 
Non-commuting (or equivalently non-Abelian) behavior is pervasive in mechanics, from the moves of a Rubik's cube
to the nonholonomic dynamics of rolling spheres and robotic arms.
Here, we focus instead on a more subtle phenomenon: the non-commutative behavior of the classical excitations (e.g., sound waves) that propagate on top of a background configuration.
The propagation of a wave packet constructed out of vibrational modes can be affected by geometric (or Berry) phases.
For a single isolated band, the Berry phases are complex numbers of modulus one that manifestly commute.
To obtain non-Abelian Berry phases, a set of (at least) two degenerate bands is required. 
The geometric phases then become $2 \times 2$ unitary matrices that need not to commute \cite{Berry1984,*Wilczek1984}.

\begin{figure*}
  \centering
  \hspace*{-1.5cm}
  \includegraphics{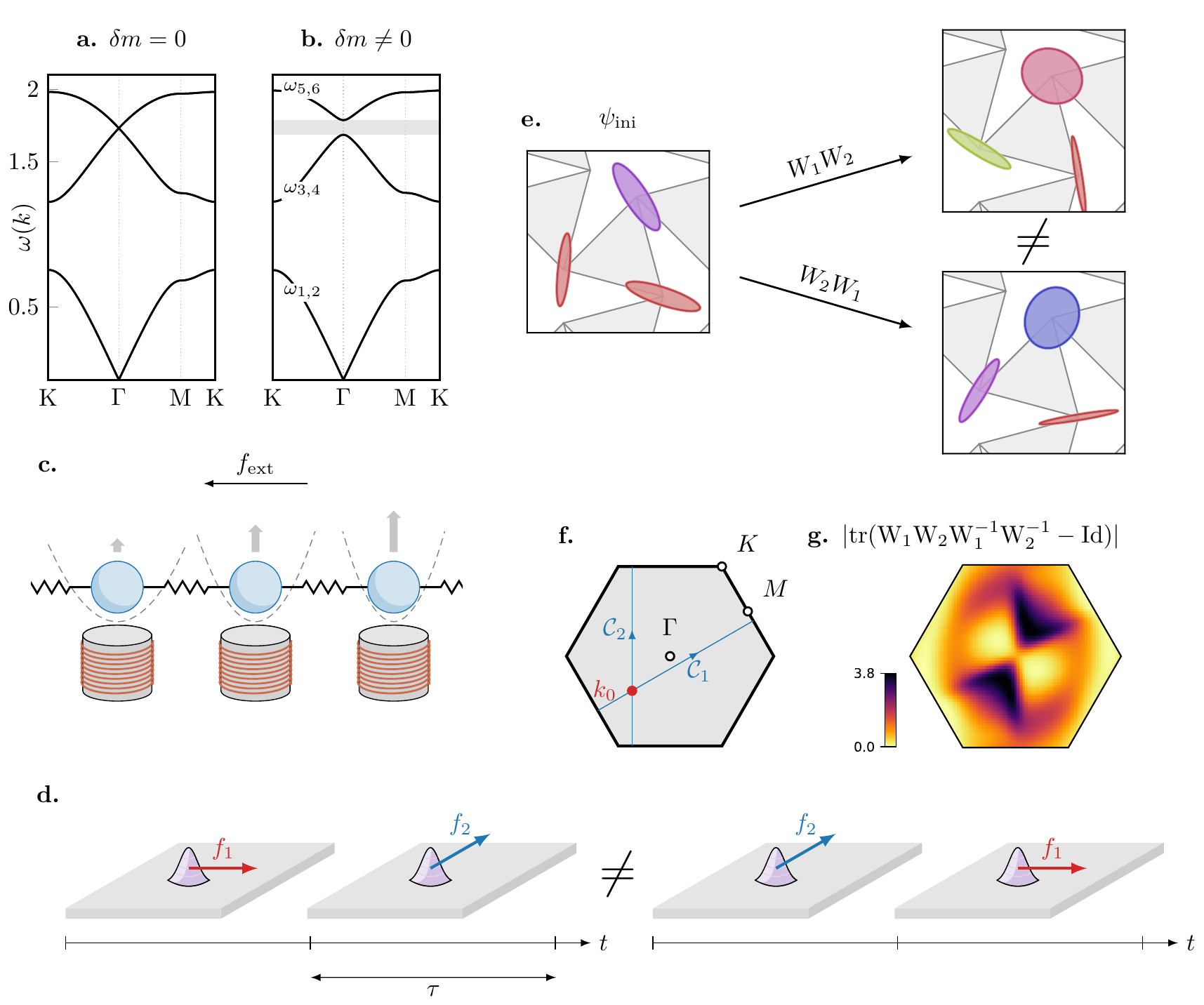}
  \caption{\label{figure_non_abelian}\strong{Non-Abelian propagation of semi-classical wave packets.}
  (a,b) The double Dirac cone becomes massive (gapped) when an asymmetry $\delta m$ is introduced between the three masses in a unit cell $(m_1,m_2,m_3) = (1-\delta m, 1, 1+\delta m)$. 
  However, the global two-fold degeneracy is preserved. 
  The gap (highlighted in gray) is proportional to $\delta m$ at first order. 
  The semi-classical propagation of a wave packet composed of vibrations with frequencies in the range of the two spectrally isolated bands with dispersions $\omega_{3}(k) = \omega_4(k)$ is affected by their geometric phases.
  The band structures are obtained by diagonalizing the Bloch dynamical matrices $D(\theta, k)$ (see \SUPPL for details).
  (c) To produce an effective force $f_{\text{ext}}$ acting on the wave packets, a spatially varying harmonic potential is superimposed to the structure. As a consequence, the dynamical matrix is modified as $D_{p q} \to D_{p q} + \delta_{p q} [\Delta \omega^2](h_q)$ where $h_q$ is the harmonic potential applied to the mass $i$. 
  For a uniform potential $h_q = h$, the optical bands are essentially shifted in frequency (see \SUPPL) by an amount proportional to $h$.
  (d) We consider a situation where sequences of constant effective forces $(f_1,f_2)$ and $(f_2,f_1)$ are applied to the system.
  The duration $\tau_n$ of each step is chosen so that the change in momentum $f_n \tau_n$ is exactly the size of a reciprocal lattice vector $|a_i^*|$ (masses are equal to one by convention).
  (e) After one sequence, the composition of a wave packet initially centered at $k_0$ in momentum space is transformed in a way described by the composition of Wilson loops, respectively $W_2 W_1$ and $W_1 W_2$, transforming a state $\varphi_{\text{ini}}$ into $\varphi_{\text{12}} = W_2 W_1 \varphi_{\text{ini}}$ or $\varphi_{\text{21}} = W_1 W_2 \varphi_{\text{ini}}$.
  The difference in the vibrational states $\varphi_{\text{21}}$ and $\varphi_{\text{12}}$ unambiguously shows the non-commutativity of the operations.
  In the picture, the Bloch vibrational states are represented by ellipses describing the motion of the masses, with a color representing their phases.
  The Wilson loop operators and their action are computed numerically for $k_0=(2,1)$ and $\delta m = \num{0.1}$. (see \SUPPL).
  (f) Brillouin zone of the triangular Bravais lattice. In blue, the trajectories in momentum space corresponding to the Wilson loops $W_{1}$ and $W_{2}$ are shown, with their base point $k_0$ in red.
  (g) A quantitative measure of the non-commutativity is obtained by comparing $W_1 W_2 W_1^{-1} W_2^{-1}$ to the identity. 
  We plot the absolute value of the trace of their difference as a function of the starting point~$k_0$ of the protocol.
  See \SUPPL for details on the numerical computation and for a discussion of the relevant orders of magnitude.
  }
\end{figure*}

The self-dual Kagome lattice is a suitable platform to realize non-Abelian sound because it has a two-fold degenerate phonon spectrum.
In order to spectrally isolate a single two-fold degenerate band, we assign different values to the three masses in the unit cell. 
As a result, the double Dirac cone at $\Gamma$ becomes gapped, see figure~\ref{figure_non_abelian}a and \ref{figure_non_abelian}b where we label the dispersion relations with increasing frequencies $\omega_{i}(k)$ with $i=1, \dots, 6$.
This modification preserves the self-dual symmetry \eqref{duality_relation}, so the global two-fold degeneracy persists regardless of the values of the masses.
Consider an acoustic wave packet constructed from the spectrally isolated central set of two-fold degenerate bands with dispersions $\omega_{3}(k) = \omega_{4}(k)$ [the same analysis could be done on the upper bands $\omega_{5}(k) = \omega_{6}(k)$].
As we apply external forces to the wave packet, it evolves, but it is constrained within the two-fold degenerate subset of mechanical vibrations as long as the external perturbation is small enough compared to the distance with the other bands (so the non-adiabatic Landau-Zener transitions can be neglected, see SI for orders of magnitudes).
The evolution of the wave packet can effectively be described by three so-called semi-classical variables: the semi-classical position $r(t)$, momentum $k(t)$, and composition (generalized polarization) $\eta(t)$ of the wave packet in the two-fold degenerate subspace \cite{Culcer2005,Shindou2005}.
The evolution of those variables is described by semi-classical equations of motion \cite{Xiao2010,Culcer2005,Shindou2005} (see also \SUPPL), originating in electronic solid-state physics, but commonly applied to other waves such as light~\cite{Onoda2004,*Onoda2006,Bliokh2007} or acoustic waves~\cite{Bliokh2006,Torabi2009,*Mehrafarin2009}.

We focus on a simple physical situation analogous to Bloch oscillations in solid-state physics, where an additional harmonic potential is imposed to each mass with a spatially-dependent stiffness as illustrated in figure~\ref{figure_non_abelian}c.
A linear increase in the stiffness pattern corresponds to a constant force on the wave packet.
Under this constant force, the momentum $k(t)$ increases linearly in time.
As the quasi-momentum $k$ is defined on the Brillouin zone, which is topologically a torus, this linear increase corresponds to a periodic evolution in time where $k(t)$ traces loops $\mathcal{C}_i$ on the Brillouin torus, as represented in figure~\ref{figure_non_abelian}f.

When an effective force is applied, the composition of the semi-classical wave packet changes from an initial polarization $\varphi_{\text{ini}}$ to $W[\mathcal{C}] \varphi_{\text{ini}}$, where $W[\mathcal{C}] = P \exp\left( - \int_{\mathcal{C}} A \right)$ is a Wilson line operator, the non-Abelian analogue of a Berry phase.
This operator is a path-ordered exponential along the path $\mathcal{C}$ in momentum space traversed under the effective force, and $A$ is the non-Abelian Berry connection \cite{Berry1984,*Wilczek1984} describing the spectrally isolated two-fold degenerate band. 
Both $A$ and $W[\mathcal{C}]$ can be directly evaluated from the normal modes of vibration obtained by diagonalizing the dynamical matrix (see \SUPPL).
For simplicity, we assume that the effective force is applied for a duration $\tau$ chosen so that the momentum changes by exactly one reciprocal lattice vector $a_i^*$, going along a closed loop from $k_0$ to $k_0 + a_i^*$ defined by $\mathcal{C}_{i}(\lambda) = k_0 + \lambda a_{i}^*$, see figure~\ref{figure_non_abelian}f.
Hence, $\mathcal{C}$ is a closed loop, and $W[\mathcal{C}]$ is called a Wilson loop operator.

After the forces $(f_1, f_2)$ are sequentially applied during the appropriate duration, as represented in figure~\ref{figure_non_abelian}d, the composition of a wave-packet initially at $k_0$ changes from any initial state $\varphi_{\text{ini}}$ to $\varphi_{\text{12}} = W_2 W_1 \varphi_{\text{ini}}$. 
The reversed sequence of mechanical actions $(f_2, f_1)$ produces a different final vibrational state $\varphi_{\text{21}} = W_1 W_2 \varphi_{\text{ini}}$, because the two Wilson loops do not commute in general,
\begin{equation}
  W_1 W_2 \neq W_2 W_1.
\end{equation}
Hence, the corresponding mechanical actions do not commute either! 
Figure~\ref{figure_non_abelian}e shows that the vibrational state of the system $\varphi_{\text{12}}$ after the application of~$f_1$ followed by~$f_2$ is different from its vibrational state $\varphi_{\text{21}}$ after the application of~$f_2$ followed by~$f_1$.
In figure~\ref{figure_non_abelian}g, we assess how the choice of the initial point $k_0$ affects the non-commutativity of $W_1(k_0)$ and $W_2(k_0)$, by quantifying the deviation of $W_1 W_2 W_1^{-1} W_2^{-1}$ from the identity. This non-commuting behaviour shares similarities with non-Abelian excitations like anyons~\cite{Stern2013,Iadecola2016,Barlas2019}. 
However, in the present study non-commutativity arises from how independent wave packets respond to external forces, while for anyons it is associated with the exchange (braiding) of these quasi-particles with each other.

%\medskip

Our results raise the prospect of materials where information is encoded and processed using non-Abelian mechanical excitations; more broadly they illustrate the power of duality relations in wave physics.
We envision that dualities and their breaking will play a key role in the design of metamaterials, as symmetries currently do.

\medskip

\clearpage
\begin{center}
\noindent \textbf{\uppercase{Supplementary Information}}
\end{center}

\setcounter{equation}{0}
\setcounter{figure}{0}

\renewcommand{\theequation}{S\arabic{equation}}
\renewcommand{\figurename}{{\bf Fig. }}
\renewcommand{\thefigure}{{\bf S\arabic{figure}}}

\makeatletter
\c@secnumdepth=4
\makeatother

\section{The dynamical matrix and the duality operator in real space}

\subsection{Dynamical matrix}

The dynamical matrix $\hat{D}$ summarizes the linearized Newton equations of motion
\begin{equation}
  \partial_t^2 \ket{\phi} = \hat{D} \ket{\phi}
\end{equation}
where $\ket{\phi}$ is a vector with components $\phi_p = \sqrt{m_p} u_p$, and where $u_p$ is the displacement of the particle $p$ with mass $m_p$ from its equilibrium position.
It describes the normal modes of vibration of a mechanical structure and their oscillation frequencies \cite{Born1954,Maradudin1971,Maradudin1968,Warren1968}.
In a system of masses coupled by springs (or more generally of coupled harmonic oscillators), it is convenient to write it as $D = M^{-1/2} Q K C M^{-1/2}$ where $M$ is a matrix containing the masses of the oscillators, $K$ a matrix containing the stiffnesses of the bonds connecting the oscillators, and $Q=C^{\dagger}$ describes the geometry and connectivity of the masses \cite{Pellegrino1986,Hutchinson2003}. 
This assumes that the mass matrix $M$ is positive-definite; this is the case in standard mechanical systems.
This version of the dynamical matrix can be seen as obtained from a canonical change of variables $(u, M \dot{u}) \mapsto (\sqrt{M} u, \sqrt{M}^{-1} M \dot{u}) = (\phi, \pi)$ in the Hamiltonian description of the system, before linearizing the equations of motions.
The corresponding linearized canonical Hamilton equations of motion read
\begin{equation}
  \partial_t \begin{pmatrix} \phi \\ \pi \end{pmatrix}
  = \begin{pmatrix} 0 & 1 \\ - \hat{D} & 0 \end{pmatrix} \begin{pmatrix} \phi \\ \pi \end{pmatrix}
\end{equation}
where $\pi = \dot{\phi}$, which can write $\partial_t \psi = \hat{L} \psi$. 
This first-order formulation is essential for the analysis of the semi-classical equations.

\medskip

In a crystal, the dynamical matrix is an operator of the form
\begin{equation}
  \hat{D} = \sum_{x,y \in \mathcal{C}} \ket{x, \mu} D_{\mu,\nu}(x,y) \bra{y, \nu}
\end{equation}
where $x$ and $y$ are points of the crystal $\mathcal{C}$, and where $\ket{x, \mu}$ represents a displacement of the mass located at $x$ along the direction $\mu$.
To take advantage of the spatial periodicity of the crystal, we decompose the crystal $\mathcal{C} = \Gamma \cdot \mathcal{F}$ into a unit cell $\mathcal{F}$ repeated along a Bravais lattice $\Gamma$, and write
\begin{equation}
  \hat{D} = \sum_{\gamma \Gamma} D(\gamma) \hat{T}_{\gamma}
\end{equation}
where $\hat{T}_{\gamma}$ is the translation operator by $\gamma$, and where $D(\gamma)$ is a matrix acting on the internal degrees of freedom in the unit cell.

\medskip

The dynamical matrix $\hat{D}(\theta)$ of the twisted Kagome lattice with twisting angle $\theta$ reads
\begin{equation}
\label{dynamical_matrix_real_space}
\begin{split}
  \hat{D}(\theta) = \begin{pmatrix}
    d_{11}(\theta) & d_{12}(\theta) & d_{13}(\theta) \\
    d_{21}(\theta) & d_{22}(\theta) & d_{23}(\theta) \\
    d_{31}(\theta) & d_{32}(\theta) & d_{33}(\theta)
  \end{pmatrix} \\
  + 
  \begin{pmatrix}
    0 & e_{12}(\theta) \, \hat{T}_{a_2} & e_{13}(\theta) \, \hat{T}_{a_1}^{-1} \\
    e_{21}(\theta) \, \hat{T}_{a_2}^{-1} & 0 & e_{23}(\theta) \, \hat{T}_{a_1}^{-1} \hat{T}_{a_2}^{-1} \\
    e_{31}(\theta) \, \hat{T}_{a_1} & e_{32}(\theta) \, \hat{T}_{a_1} \hat{T}_{a_2} & 0 
  \end{pmatrix}
\end{split}
\end{equation}
where $\hat{T}_{\gamma}$ is the translation operator by $\gamma$, satisfying $\hat{T}_{\gamma}^\dagger = \hat{T}_{-\gamma} = \hat{T}_{\gamma}^{-1}$ and $\hat{T}_{\gamma} \hat{T}_{\rho} = \hat{T}_{\gamma + \rho} = \hat{T}_{\rho} \hat{T}_{\gamma}$.
(The identity in position space $\hat{T}_{0}$ is implied in the first matrix in equation \eqref{dynamical_matrix_real_space}.)
We have written $\hat{D}$ as a matrix acting on the masses,
meaning that the elements $d_{mn}$ and $e_{mn}$ map the mass $M_n$ to the mass $M_m$.
Besides, $d_{mn}$ and $e_{mn}$ are also $2 \times 2$ matrices acting on the $x$ and $y$ components of the displacements of the masses.
We have chosen the dynamical matrix to be Hermitian, so $d_{m n}^\dagger = d_{n m}$, and $e_{m n}^\dagger = e_{n m}$.
The blocks are
% generated_by(symbolic_dynamical_matrix.ipynb)
\begin{subequations}
\begin{align}
d_{11}(\theta) &= \begin{pmatrix}\frac{\cos{\left (2 \theta \right )}}{2} + 2 & - \frac{\sqrt{3} \cos{\left (2 \theta \right )}}{2}\\- \frac{\sqrt{3} \cos{\left (2 \theta \right )}}{2} & - \frac{\cos{\left (2 \theta \right )}}{2} + 2\end{pmatrix}\\
d_{22}(\theta) &= \begin{pmatrix}- \cos{\left (2 \theta \right )} + 2 & 0\\0 & \cos{\left (2 \theta \right )} + 2\end{pmatrix}\\
d_{33}(\theta) &= \begin{pmatrix}\frac{\cos{\left (2 \theta \right )}}{2} + 2 & \frac{\sqrt{3} \cos{\left (2 \theta \right )}}{2}\\\frac{\sqrt{3} \cos{\left (2 \theta \right )}}{2} & - \frac{\cos{\left (2 \theta \right )}}{2} + 2\end{pmatrix}\\
d_{12}(\theta) &= \begin{pmatrix}\frac{\cos{\left (2 \theta + \frac{\pi}{3} \right )}}{2} - \frac{1}{2} & \frac{\sin{\left (2 \theta + \frac{\pi}{3} \right )}}{2}\\\frac{\sin{\left (2 \theta + \frac{\pi}{3} \right )}}{2} & - \frac{\cos{\left (2 \theta + \frac{\pi}{3} \right )}}{2} - \frac{1}{2}\end{pmatrix}\\
d_{13}(\theta) &= \begin{pmatrix}- \cos^{2}{\left (\theta \right )} & - \frac{\sin{\left (2 \theta \right )}}{2}\\- \frac{\sin{\left (2 \theta \right )}}{2} & - \sin^{2}{\left (\theta \right )}\end{pmatrix}\\
d_{23}(\theta) &= \begin{pmatrix}\frac{\sin{\left (2 \theta + \frac{\pi}{6} \right )}}{2} - \frac{1}{2} & - \frac{\cos{\left (2 \theta + \frac{\pi}{6} \right )}}{2}\\- \frac{\cos{\left (2 \theta + \frac{\pi}{6} \right )}}{2} & - \frac{\sin{\left (2 \theta + \frac{\pi}{6} \right )}}{2} - \frac{1}{2}\end{pmatrix}\\
e_{12}(\theta) &= \begin{pmatrix}\frac{\sin{\left (2 \theta + \frac{\pi}{6} \right )}}{2} - \frac{1}{2} & \frac{\cos{\left (2 \theta + \frac{\pi}{6} \right )}}{2}\\\frac{\cos{\left (2 \theta + \frac{\pi}{6} \right )}}{2} & - \frac{\sin{\left (2 \theta + \frac{\pi}{6} \right )}}{2} - \frac{1}{2}\end{pmatrix}\\
e_{13}(\theta) &= \begin{pmatrix}- \cos^{2}{\left (\theta \right )} & \frac{\sin{\left (2 \theta \right )}}{2}\\\frac{\sin{\left (2 \theta \right )}}{2} & - \sin^{2}{\left (\theta \right )}\end{pmatrix}\\
e_{23}(\theta) &= \begin{pmatrix}\frac{\cos{\left (2 \theta + \frac{\pi}{3} \right )}}{2} - \frac{1}{2} & - \frac{\sin{\left (2 \theta + \frac{\pi}{3} \right )}}{2}\\- \frac{\sin{\left (2 \theta + \frac{\pi}{3} \right )}}{2} & - \frac{\cos{\left (2 \theta + \frac{\pi}{3} \right )}}{2} - \frac{1}{2}\end{pmatrix}.
\end{align}
\end{subequations}

\subsection{Duality}

Let us consider the \ang{90} rotation matrix in the displacement spaces,
\begin{equation}
  r_{\blsq} = \ii \sigma_y = \begin{pmatrix}
    0 & 1 \\
    - 1 & 0
  \end{pmatrix}
\end{equation}
satisfying $r_{\blsq}^2 = - \Id$.
We observe that 
\begin{subequations}
\label{duality_relations_on_blocks}
\begin{align}
  r_{\blsq} \, d_{m n}(\theta) \, r_{\blsq}^{-1} &= e_{m n}(\theta^*) \quad \text{for $m \neq n$,} \\
  r_{\blsq} \, d_{m m}(\theta) \, r_{\blsq}^{-1} &= d_{m m}(\theta^*)
\end{align}
\end{subequations}
where $\theta^* = 2 \theta_{\text{c}} - \theta$ and $\theta_{\text{c}} = \pi/4$.

With $R_{\blsq} = \text{diag}(r_{\blsq}, r_{\blsq}, r_{\blsq})$, this implies
\begin{equation}
\begin{split}
  R_{\blsq} \hat{D}(\theta) R_{\blsq}^{-1} = \begin{pmatrix}
    d_{11}(\theta^*) & e_{12}(\theta^*) & e_{13}(\theta^*) \\
    e_{21}(\theta^*) & d_{22}(\theta^*) & e_{23}(\theta^*) \\
    e_{31}(\theta^*) & e_{32}(\theta^*) & d_{33}(\theta^*)
  \end{pmatrix} \\
  + 
  \begin{pmatrix}
    0 & d_{12}(\theta^*) \hat{T}_{a_2} & d_{13}(\theta^*) \hat{T}_{a_1}^{-1} \\
    d_{21}(\theta^*) \hat{T}_{a_2}^{-1} & 0 & d_{23}(\theta^*) \hat{T}_{a_1}^{-1} \hat{T}_{a_2}^{-1} \\
    d_{31}(\theta^*) \hat{T}_{a_1} & d_{32}(\theta^*) \hat{T}_{a_1} \hat{T}_{a_2} & 0 
  \end{pmatrix}.
\end{split}
\end{equation}

Hence, let us define a unitary operator $\hat{\mathcal{V}}$ combining $R_{\blsq}$ and lattice translations as
\begin{equation}
  \label{translations_for_duality}
  \hat{\mathcal{V}} = \begin{pmatrix}
    r_{\blsq} \, \hat{T}_{\gamma_0} & 0 & 0 \\
    0 & r_{\blsq} \, \hat{T}_{\gamma_0+a_2} & 0 \\
    0 & 0 & r_{\blsq} \, \hat{T}_{\gamma_0-a_1}
  \end{pmatrix}
\end{equation}
Here, $\gamma_0$ is an arbitrary reciprocal lattice vector that we will set to zero.

From equation \eqref{duality_relations_on_blocks}, we can see that
\begin{equation}
\begin{split}
  \hat{\mathcal{V}} \hat{D}(\theta) \hat{\mathcal{V}}^{-1} = 
  \begin{pmatrix}
    0 & d_{12}(\theta^*) & d_{13}(\theta^*)  \\
    d_{21}(\theta^*) & 0 & d_{23}(\theta^*) \\
    d_{31}(\theta^*) & d_{32}(\theta^*) & 0 
  \end{pmatrix} \\
  + 
  \begin{pmatrix}
    d_{11}(\theta^*) & e_{12}(\theta^*) \hat{T}_{a_2}^{-1} & e_{13}(\theta^*) \hat{T}_{a_1} \\
    e_{21}(\theta^*) \hat{T}_{a_2} & d_{22}(\theta^*) & e_{23}(\theta^*) \hat{T}_{a_1} \hat{T}_{a_2} \\
    e_{31}(\theta^*) \hat{T}_{a_1}^{-1} & e_{32}(\theta^*) \hat{T}_{a_1}^{-1} \hat{T}_{a_2}^{-1} & d_{33}(\theta^*)
  \end{pmatrix}.
\end{split}
\end{equation}

We then define a unitary operator $\hat{\mathcal{I}}$ such that $\hat{\mathcal{I}} \hat{T}(\gamma) \hat{\mathcal{I}}^{-1} = \hat{T}(-\gamma)$. 
Formally, $\hat{\mathcal{I}}$ is defined by its action on the basis vectors by
\begin{equation}
  \hat{\mathcal{I}} \ket{\gamma, x_i} = \ket{-\gamma, x_i}
\end{equation}
for $\gamma \in \Gamma$ and $x_i \in \mathcal{F}$.
This operation resembles spatial inversion, but only acts on the Bravais lattice.
The definition above depends on the choice of the fundamental domain $\mathcal{F}$ used to define the basis vectors of the vibration space.
This choice does not appear to be easily avoidable due to the fact that the duality maps a given system to another system.
However, this arbitrary in the definition of $\hat{\mathcal{I}}$ appears to be inessential. 
Combining $\hat{\mathcal{I}}$ with $\hat{\mathcal{V}}$ into
\begin{equation}
  \label{unitary_duality_operator_real_space}
  \hat{\mathscr{U}} = \hat{\mathcal{I}} \hat{\mathcal{V}}
\end{equation}
we obtain the duality operator in real space, satisfying
\begin{equation}
  \label{unitary_duality_real_space}
  \hat{\mathscr{U}} \hat{D}(\theta) \hat{\mathscr{U}}^{-1} = \hat{D}(\theta^*).
\end{equation}
One can verify that $\hat{\mathscr{U}}^2 = - \Id$.

In principle, one could choose to redefine $\hat{\mathscr{U}} \to \pm \ii \hat{\mathscr{U}}$ such that it squares to $+\Id$. 
(This does not affect the property $\hat{\mathscr{A}}^2 = - \Id$ because $\hat{\mathscr{A}}$ is anti-unitary.)
However the new $\hat{\mathscr{U}}$ would map real-valued displacements to complex-valued ones.
This motivates our choice leading to $\hat{\mathscr{U}}^2 = - \Id$. 
With the alternative choice of a duality operator squaring to $+\Id$, the group-theoretical analysis of section \ref{app_symmetries_gamma} would be modified.

\section{Dualities in momentum space}

The momentum-space equivalent of \eqref{unitary_duality_real_space} reads
\begin{equation}
  \label{unitary_duality}
  \mathscr{U}(-k, k) D(\theta, k) \mathscr{U}^{-1}(-k, k) = D(\theta^*, -k) 
\end{equation}
where
\begin{equation}
  \label{duality_operator}
  \mathscr{U}(k) \equiv \mathscr{U}(k, -k) = \begin{pmatrix}
    r_{\blsq} & 0 & 0 \\
    0 & r_{\blsq} \, \ee^{-\ii k \cdot a_2} & 0 \\
    0 & 0 & r_{\blsq} \, \ee^{\ii k \cdot a_1}
  \end{pmatrix}.
\end{equation}
This equation holds with the Bloch convention where $\hat{T}_{\gamma}$ corresponds to $\ee^{\ii k \cdot \gamma}$.
In the following, we will also write $\kappa_{i} = k \cdot a_i$ to shorten the notations.

The duality $\hat{\mathscr{U}}$ maps $k$ to $-k$ (and conversely), as implied by the notation $\mathscr{U}(-k, k)$.
More precisely, it is a linear operator acting on the vector bundle of Bloch eigenmodes, mapping the fiber over~$k$ to the fiber over~$-k$.

One can verify by a direct computation that $\mathscr{U}(-k,k) \mathscr{U}(k,-k) = - \Id_{6}$, corresponding to $\hat{\mathscr{U}}^2 = - \Id$ (where $\Id$ is the fiberwise identity operator). 

In momentum space, the anti-unitary operator $\hat{\Theta} = \mathcal{K}$ (complex conjugation) also maps $k$ to $-k$, because $\mathcal{K} \ee^{-\ii k \cdot x} = \ee^{\ii k \cdot x}$. As the real-space dynamical matrices have real entries, we have
\begin{equation}
  \label{bosonic_time_reversal}
  \Theta D(\theta, -k) \Theta^{-1} = D(\theta, k).
\end{equation}
This constraint is usually called a bosonic time-reversal symmetry, although it does not correspond to classical time-reversal invariance.
Hence, the combination $\hat{\mathscr{A}} = \hat{\mathscr{U}} \hat{\Theta}$ acts anti-linearly fiberwise: it is anti-unitary, but maps each momentum $k$ to itself. 
Combining equations \eqref{unitary_duality} and \eqref{bosonic_time_reversal}, we obtain the duality
\begin{equation}
  \mathscr{A}(k) D(\theta^*, k) \mathscr{A}^{-1}(k) = D(\theta, k)
\end{equation}
where $\mathscr{A}(k) = \mathscr{U}(k) \Theta$.
One can verify that $\hat{\mathscr{A}}^2 = - \Id$ (because $\mathscr{U}(k) \overline{\mathscr{U}(k)} = - \Id$).

\medskip

In this paragraph, we have used a standard convention for the Fourier transform where the duality is easily expressed.
However, the semi-classical equations are more conveniently expressed with another convention where the families of Bloch matrices do not depend on the unit cell, as discussed in section \ref{app_bloch_conventions}.
In this alternative convention, the equivalent of the quantity $\mathscr{U}(k)$ defined in equation~\eqref{duality_operator} is
\begin{equation}
  \mathscr{U}_{\text{can}}(k) = \begin{pmatrix}
    r_{\blsq} & 0 & 0 \\
    0 & \ee^{-\ii \displaystyle\frac{(2 \kappa_{1} + \kappa_{2})}{\sqrt{3}}} r_{\blsq} & 0 \\
    0 & 0 & \ee^{-\ii \displaystyle\frac{(\kappa_{1} + 2 \kappa_{2})}{\sqrt{3}}} r_{\blsq}
  \end{pmatrix}.
\end{equation}

\section{Bloch conventions}
\label{app_bloch_conventions}

The Bloch decomposition of a spatially periodic operator can be performed in different ways.
There are at least two usual conventions for the Bloch decomposition differing in whether the phase factor attributed to translations is computed from (I) the Bravais lattice translations or (II) the crystal translations. 
We refer the reader to references \cite{Blount1962,Zak1967,Panati2003,Bena2009,Fruchart2014,Lim2015,Dobardi2015} for details.
Both conventions are useful in different situations. 
In particular, the convention (I) is the most natural when writing the semi-classical equations of motion \cite{Fruchart2014,Lim2015}. 
Reference \cite{Dobardi2015} discusses the relations with crystal symmetries.
For a given choice of fundamental domain $(x_i)_{i=1,\dots,F}$ of the crystal (here, the positions of the masses), the change of basis matrix relating both conventions is
$V_{\mathcal{F}}(k) = \diag[ ( \ee^{-\ii k (x_i-x_0)} \sigma_0 )_{i=1,\dots,F} ]$ where $x_0$ is an arbitrary origin.
% VZB st VZB DB VZB^{-1} = DZ % Z=II while B=I

\section{Symmetries and degeneracies at the $\Gamma$ point}
\label{app_symmetries_gamma}

In addition to the apparition of mechanical Kramers pairs at each point, the self-duality has interesting consequences in terms of band degeneracies. 
In this section, we discuss the interplay of symmetries and degeneracies at the center of the Brillouin zone, called the $\Gamma$ point.
Unless otherwise specified, we assume that the three masses in the unit cell are identical (similar for the springs) in order to preserve all spatial symmetries.

\subsection{Double Dirac cones}

For a generic twisting angle $\theta$, a two-fold degeneracy at finite frequency is observed at $\Gamma$, and the band crossing is quadratic. 
At the critical twisting angle, the band crossing becomes linear, see figure~\ref{figure_degeneracies_gamma}.

This is uncommon: band crossings at the center of the Brillouin zone are normally quadratic \cite{Sakoda2011,Huang2011}. More precisely, a single Dirac point at $\Gamma$ is forbidden by time-reversal invariance. This can be understood as follows: let us consider a two-bands effective Hamiltonian describing a Dirac cone around a high symmetry point $k_0 = - k_0$ (mod. reciprocal lattice vectors),
\begin{equation}
   H(q) = v_x q_x \sigma_x + v_y q_y \sigma_y
\end{equation}
where $q = k - k_0$ and $v_{x,y}$ are parameters. Let us further assume that this effective Hamiltonian is Hermitian, i.e. $H(q) = H^\dagger(q)$, meaning that $v_x$ and $v_y$ are real; and that it is time-reversal invariant in the sense that $\overline{H}(q) = H(-q)$. Hence, we find that
\begin{equation}
  v_x q_x \sigma_x - v_y q_y \sigma_y = -v_x q_x \sigma_x - v_y q_y \sigma_y
\end{equation}
This is only possible if $v_x = 0$, so the effective Hamiltonian does not describe a Dirac cone. (The same argument applies if a mass term $m \sigma_z$ is added to the effective Hamiltonian. The conclusion does not depend on the choice of parametrization, because only $\overline{\sigma_y} = - \sigma_y$.)
For this reason, finite frequency linear dispersions at the center of the Brillouin zone are unusual.

Usual approaches to obtain them rely on accidental degeneracies, where either (a) a three-fold degeneracy with an almost flat band \cite{Huang2011} or (b) two two-fold degeneracies, typically with different group velocities \cite{Sakoda2012a,Sakoda2012b,Dai2017,Li2014}, are made degenerate by tuning the structure, allowing linear dispersions. Alternatively, genuine Dirac dispersions can be obtained by directly breaking time-reversal invariance~\cite{Koutserimpas2018,Zhou2018}.

The double Dirac cone observed here can be understood as the overlap of an essential 2-fold degeneracy with two other bands, this overlap being enforced through the Kramers theorem by the anti-unitary $\mathscr{A}(\Gamma)$, guaranteeing that the group velocities of the two Dirac cones are equal.
Unlike a single Dirac cone at $\Gamma$, this situation is allowed under the conditions stated above.
This can be seen from a $k\!\cdot\!p$-like expansion of the dynamical matrix (see section~\ref{app_effective_dynamical_matrix} for details).
The matrix $D(\Gamma)$ has dimensionless eigenvalues $(0,0,3,3,3,3)$. 
The $4 \times 4$ effective dynamical matrix corresponding to the four degenerate eigenvalues involved in the double Dirac cone is, up to a constant term proportional to the identity and a multiplicative constant,
\begin{equation}
  D_{\text{eff}}(k) \simeq \begin{pmatrix}
    0 & 0 &   \ii k_y & - \ii k_x \\
    0 & 0 & - \ii k_x & - \ii k_y \\
    - \ii k_y &  \ii k_x & 0 & 0 \\
      \ii k_x &  \ii k_y & 0 & 0 
  \end{pmatrix}.
\end{equation}
where $A_{1}$ is a constant.
It satisfies both $D_{\text{eff}}(k) = D_{\text{eff}}^\dagger(k)$ $\overline{D_{\text{eff}}}(k) = D_{\text{eff}}(-k)$.
A unitary rotation brings the effective dynamical matrix in the form
\begin{equation}
  D_{\text{eff}}(k) \simeq \begin{pmatrix}
    k \cdot \sigma & 0 \\
    0 & - k \cdot \sigma
  \end{pmatrix}
\end{equation}
where $k \cdot \sigma = k_x \sigma_x + k_y \sigma_y$ and $\sigma_i$ are Pauli matrices.
In this form, it is clear that $D_{\text{eff}}(k)$ describes two superimposed Dirac cones.

\subsection{Spatial symmetries and extended symmetries at Gamma}
\label{app_extended_symmetries_gamma}

\begin{figure}
  \centering
  \includegraphics{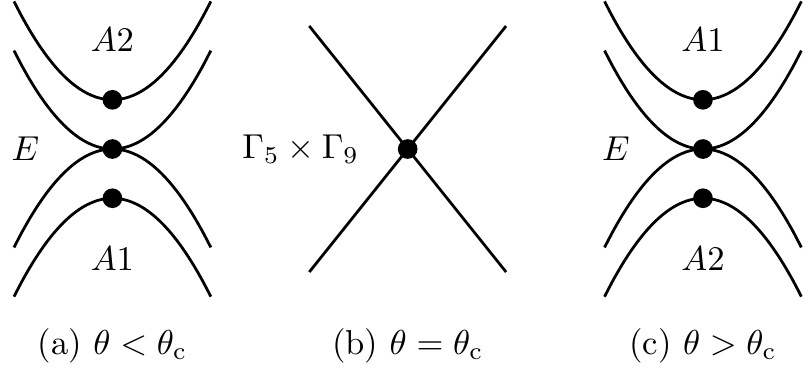}
  \caption{\label{figure_degeneracies_gamma}\strong{Degeneracies at the center of the Brillouin zone.}
  }
\end{figure}

From a group-theoretical perspective, the unitary self-duality relation implies that the group of symmetries of the dynamical matrix is enhanced. In particular, the point group at time-reversal invariant momenta $k^*$ such that $k^* = -k^*$ modulo a reciprocal lattice vector effectively acquires a new generator, because
\begin{equation}
  \mathscr{U}(k^*) D(k^*) \mathscr{U}^{-1}(k^*) = D(k^*).
\end{equation}

With the aim of better understanding the nature of the double Dirac cone discussed in the previous section, we focus on the $\Gamma$ point ($k=0$), where the unitary $\mathscr{U}_0 = \mathscr{U}(\Gamma)$ is a symmetry of the system. 
The point group of crystal symmetries (little co-group) at $\Gamma$ (not taking $\mathscr{U}_0$ into account) is always \sch{C3v} (\hm{3m} in Hermann-Mauguin notation), whether the Kagome lattice is critical or not.
(The point groups at $\Gamma$ are directly obtained from the plane groups (2D space groups) determined in section~\ref{app_sec_kagome_plane_groups}, and the plane group of the structure does not change at the critical angle.)

This point group has to be supplemented with $\mathscr{U}_0$.
Hence, we consider the matrix group obtained by combining the generators of (the representation acting on the dynamical matrix of) \sch{C3v} with $\mathscr{U}_0$.
The matrix group was identified with the computational group theory software GAP \cite{GAP4} to be isomorphic to $D_{12}$, the dihedral of order \num{24} (we follow the crystallographic convention where the dihedral group of order $2n$ is called $D_{n}$, and not the abstract algebra convention where the same group is called $D_{2n}$), which can alternatively be seen as \sch{C12v} (the point groups \sch{D12}, \sch{C12v}, and \sch{D6d} are isomorphic).
Interestingly, while $D_{12}$ is a two-dimensional point group, it is not a crystallographic point group in 3D (nor in 2D) as it has a $12$-fold axis (obtained by combining $\mathscr{U}_0$ with a $3$-fold rotation of \sch{C3v}), incompatible with the translation symmetry of a 3D (or 2D) crystal (due to the crystallographic restriction theorem, see e.g.~\cite{Senechal1996}).
This shows that non-crystallographic symmetries emerge as a consequence of the self-duality of the critical lattice. 

The representatives of conjugacy classes (used to compute the characters of our representations) as well as the character table of the group, provided in table \ref{character_table_D12}, were also computed using GAP \cite{GAP4}.
Although the existence of non-crystallographic symmetries in the symmetry group at $\Gamma$ is a surprising feature, this extended symmetry does not appear to be the origin of the four-fold degeneracy (the double Dirac cone). 
In particular, the inspection of the irreducible representations of $D_{12}$ (see the character table \ref{character_table_D12}) shows that they are at most $2$-dimensional. 
Furthermore, their 2$^\text{nd}$ Frobenius-Schur indicators (computed with GAP) are all equal to~\num{1}, meaning that the irreducible representations are real (so they should not be combined into larger-dimensional real representations). 
Hence, there is no essential four-fold degeneracy at $\Gamma$, in the sense that there is no underlying 4-dimensional irreducible representation.
However, the overlap of the two 2D IR is indeed not accidental, in the sense that it enforced through the Kramers theorem by the anti-unitary $\mathscr{A}(\Gamma)$.

\begin{figure}
  \centering
  \includegraphics{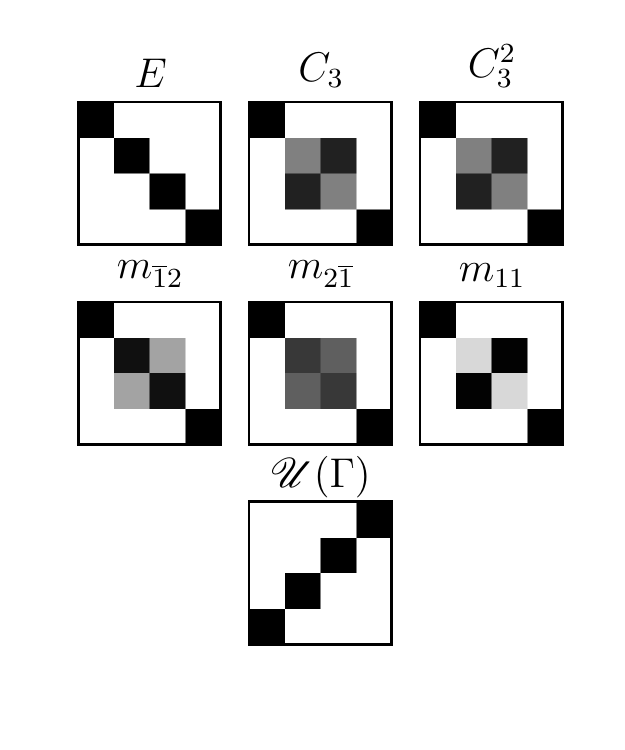}
  \caption{\label{C3v_blocks_class_representatives_and_U}\strong{Block-diagonalization of the spatial symmetries.}
  The spatial symmetries at $\Gamma$ (forming the point group \sch{C3v}) are numerically block-diagonalized in a common basis.
  We plot the absolute values of the matrix elements of class representatives of \sch{C3v}, and of the self-dual symmetry $\mathscr{U}_0$.
  This procedure yields three blocks of sizes $(1,2,1)$ corresponding to the decomposition $A_1 \oplus E \oplus A_2$ of \sch{C3v} expected from group theory.
  However, the self-dual symmetry $\mathscr{U}_0$ at $\Gamma$ is not block-diagonal in this basis.
  }
\end{figure}

\begin{figure}
  \centering
  \includegraphics{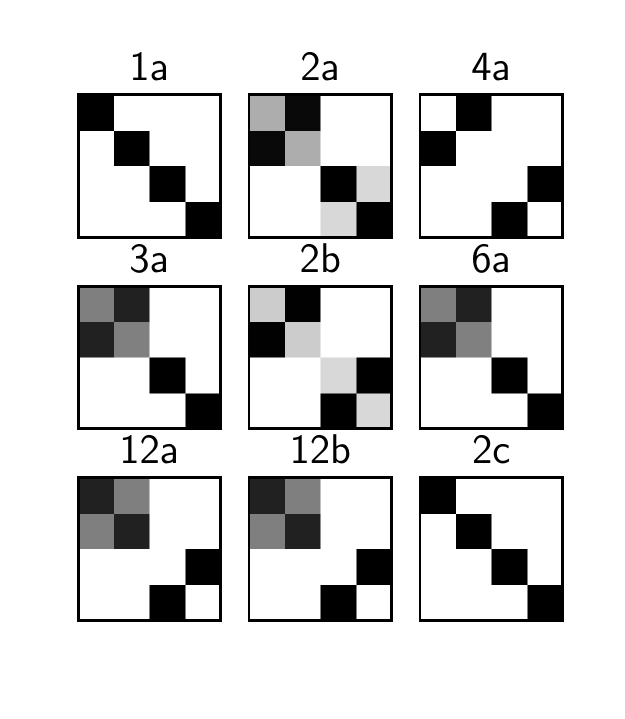}
  \caption{\label{D12_blocks_class_representatives}\strong{Block-diagonalization of the spatial and self-dual symmetries.}
  When including the self-dual symmetry, the block-diagonalization produces a different result, with two blocks of sizes $(2,2)$.
  This corresponds to the decomposition $\Gamma_{5} \oplus \Gamma_{9}$ of the enhanced symmetry group $D_{12}$.
  Here, we plotted the matrix elements of the class representatives of $D_{12}$ (corresponding to the columns in the character table \ref{character_table_D12}) obtained by combining the spatial symmetries and $\mathscr{U}_0$.
  }
\end{figure}

We now further analyze the irreducible representations at $\Gamma$.
It is instructive to first extend the analysis to non-critical lattices (with $\phi \equiv \theta - \theta_{\text{c}} \neq 0$) to understand the evolution of the degeneracies when the twisting angle crosses its critical value, see figure~\ref{figure_degeneracies_gamma}.
Outside of the critical point (where $\mathscr{U}_0$ is not a symmetry), the six bands at the $\Gamma$ point correspond, for increasing frequencies, to the irreducible representations $E$ (at zero frequency), $A_1$, $E$, and $A_2$ of \sch{C3v} for $\theta < \theta_{\text{c}}$, and $A_{1/2}$ are exchanged for $\theta > \theta_{\text{c}}$. 
Let us first only consider \sch{C3v} symmetries. 
At the critical point, we can extrapolate this picture as follows: the anti-unitary $\mathscr{A}_0$ enforces a degeneracy between the bands in the 2D IR $E$ at $\Gamma$ on one side, and the bands which end up in the 1D IRs $A_1$ and $A_2$ on the other side. 
(The bosonic time-reversal is never broken, so $E$ can be understood as a real 2D IR.) 
Hence, we can expect that the 4-fold degeneracy at $\phi = 0$ can be decomposed as $A_1 \otimes E \otimes A_2$.
While this is true, this decomposition is not entirely meaningful as it ignores the additional symmetry $\mathscr{U}_0$, that does not preserve this decomposition.
When $\mathscr{U}_0$ is taken into account, the 4-fold degeneracy can instead be seen as the product $\Gamma_{5} \times \Gamma_{9}$ of two 2D IR of $D_{12}$.
This can be seen by an explicit analysis of the symmetry operators.
To do so, we use the algorithm of reference \cite{Maehara2011} to simultaneously block-diagonalize all the symmetry operators in a common basis.
We first ignore $\mathscr{U}_0$ and block-diagonalize all other symmetries. 
We find that \sch{C3v} symmetries are indeed block-diagonalized in a common basis, with two $1 \times 1$ blocks (for $A_1$ and $A_2$) and one $2 \times 2$ block (for $E$).
However, this is not the case of the self-dual symmetry $\mathscr{U}_0$ that preserves the $E$ block but exchanges the blocks $A_1$ and $A_2$.
This can be seen from figure \ref{C3v_blocks_class_representatives_and_U} where we plot the absolute value of the matrix elements for a representative for each conjugacy class of \sch{C3v} for the decomposition $A_1 \oplus E \oplus A_2$, as well as the matrix elements of $\mathscr{U}_0$ in the same basis.
Block-diagonalizing all symmetries including $\mathscr{U}_0$ indeed leads to two $2 \times 2$ blocks corresponding to the IR $\Gamma_{5}$ and $\Gamma_{9}$ of the enhanced symmetry group $D_{12}$.
In figure \ref{D12_blocks_class_representatives}, we plot the absolute value of the matrix elements of the conjugacy class representatives of $D_{12}$ for the decomposition $\Gamma_{5} \times \Gamma_{9}$, where it is apparent that all symmetries including the ones constructed from $\mathscr{U}_0$ are block-diagonal.

\begin{table*}[htb]
\centering
\sisetup{table-format=1.0,table-column-width=1cm}
\begin{tabular}{lS[]S[]S[]S[]S[]S[]S[]S[]S[]}
\toprule
\sch{D12} & 1 & 2 & 4 & 3 & 2' & 6 & 12 & 12' & 2''\\
mult. &1 & 6 & 2 & 2 & 6 & 2 & 2 & 2 & 1\\
\midrule
$\Gamma_1$ $\;$ \irrep{A1} & 1 & 1 & 1 & 1 & 1 & 1 & 1 & 1 & 1\\
$\Gamma_2$ $\;$ \irrep{A2} & 1 & -1 & 1 & 1 & -1 & 1 & 1 & 1 & 1\\
$\Gamma_3$ $\;$ \irrep{B1} & 1 & -1 & -1 & 1 & 1 & 1 & -1 & -1 & 1\\
$\Gamma_4$ $\;$ \irrep{B2} & 1 & 1 & -1 & 1 & -1 & 1 & -1 & -1 & 1\\
$\Gamma_5$ $\;$ \irrep{E3} & 2 & 0 & 0 & 2 & 0 & -2 & 0 & 0 & -2\\
$\Gamma_6$ $\;$ \irrep{E2} & 2 & 0 & -2 & -1 & 0 & -1 & 1 & 1 & 2\\
$\Gamma_7$ $\;$ \irrep{E4} & 2 & 0 & 2 & -1 & 0 & -1 & -1 & -1 & 2\\
$\Gamma_8$ $\;$ \irrep{E1} & 2 & 0 & 0 & -1 & 0 & 1 & {$\sqrt{3}$} & {$-\sqrt{3}$} & -2\\
$\Gamma_9$ $\;$ \irrep{E5} & 2 & 0 & 0 & -1 & 0 & 1 & {$-\sqrt{3}$} & {$\sqrt{3}$} & -2\\
\bottomrule
\end{tabular}
\caption{\label{character_table_D12}\strong{Character table of the dihedral group of order 24.} 
Character table of the dihedral group of order~\num{24},~\sch{D12}. The character table was obtained in GAP \cite{GAP4} from the matrix group generated by the representation of \sch{C3v} acting on $D(\Gamma)$ supplemented with the matrix $\mathscr{U}_0$. 
We also indicate Mulliken symbols \cite{Mulliken1955} for the irreducible representations, but their attribution is partially arbitrary as it depends on choices of the axes, etc.
}
\end{table*}

\section{Mechanical Bloch oscillations}
\label{app_mechanical_bloch_oscillations}

In the main text, we consider mechanical Bloch oscillations, where the external force field appearing acting on the wave packets in the semi-classical equations is uniform.
In this section, we write the simplified semi-classical equations of motion in this situation and discuss a possible implementation of the external potential (see SI and references \cite{Culcer2005,Onoda2006,Bliokh2007,Bliokh2006} for a more detailed discussion of the semi-classical equations).

To do so, we first consider that the effective Hamiltonian $\mathcal{H}(r,k) = \Omega(k) + V(r)$ appearing in the semi-classical equations is a scalar matrix (proportional to the identity), and where the external potential $V$ does not depend on $k$ in first approximation.
In this case, the semiclassical equations \eqref{semiclassical_eom} have the simpler form
\begin{equation}
\begin{split}
  \dot{r}^{\mu} &=  \frac{\partial \Omega}{\partial k_{\mu}} + \ii \braket{F^{\mu \nu}}_{\eta} \, \dot{k}_{\nu}  \\
  \dot{k}_{\mu} &= - \frac{\partial V}{\partial r^{\mu}} \\
  \dot{\eta}   &= - (\ii \mathcal{H} + A^{\mu} \dot{k}_{\mu}) \, \eta
\end{split}
\end{equation}
The case of a uniform force $f_0 = - \partial_r V$ described in the main text describes so-called Bloch oscillations.
In this case, then the momentum equation is trivial as $\dot{k} = f_0$ is also constant, so $k(t) = k(0) + f_0 t$ and we can solve the composition equation up to a $U(1)$ phase, namely $\eta(t) = \ee^{\ii \phi(t)} W[k](t) \eta(0)$ where $W[\mathcal{C}]$ is the Wilson line operator associated to the trajectory $t \mapsto \mathcal{C}(t)$ in momentum space, $W[\mathcal{C}] = P \exp\left( - \int_{\mathcal{C}} A \right)$ and where $\phi(t)$ is an unknown phase which depends on the trajectory in real space. 
The Wilson line operator can usually not be obtained analytically, but it can be computed numerically, see section \ref{app_wilson_loops}.
The equation in real space does not depend on $\phi(t)$, and can then be solved independently.

\begin{figure}
  \centering
  \includegraphics{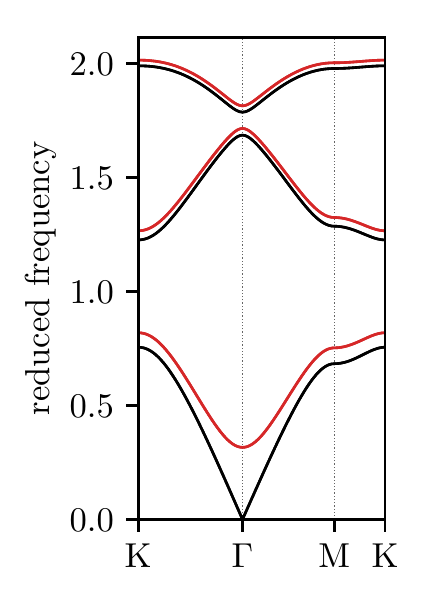}
  \caption{\label{effect_of_harmonic_potential}\strong{Effect of an additional harmonic potential.}
  The reduced frequency bands $\omega_i(k)/\omega_0$ are plotted on a high-symmetry path on the Brillouin zone, 
  in black for $(\Delta \omega / \omega_0)^2 = \num{0}$ (without additional potentials)
  and in red for $(\Delta \omega / \omega_0)^2 = \num{0.1}$ (with additional potentials).
  Here, we have set $\delta m = \num{0.1}$.
  }
\end{figure}

In practice, an external potential leading to a constant effective force can be implemented by patterning the system with a additional harmonic potential applied to each mass.
In this case, the dynamical matrix is modified from $D_{m n}$ $D_{m n} + \delta_{m n} \Delta \omega_i^2$ where $\Delta \omega_i$ is the characteristic frequency associated to the stiffness of the additional potential.
Assuming that the spatial variation of $\Delta \omega_i$ is slow (as required for the validity of the semi-classical equations), we can understand the effects of this modification of the system by looking at the change in the band structure in a system where $\Delta \omega_i$ is uniform.
They are illustrated in figure \ref{effect_of_harmonic_potential}.
As expected, those effect are drastic for the zero-modes of the system which are immediately lifted but the optical bands we are interested in here are essentially globally shifted in frequency. 

A similar scheme was already experimentally realized (although without spatial variations) in lattices of gyroscopes \cite{Mitchell2018,Nash2015} to control topological phase transitions.

\section{Wilson line operators and their numerical computation}
\label{app_wilson_loops}

In this section, we recall standard results about Wilson line and loop operators \cite{Wilson1974,*Berry1984,*Wilczek1984} and their numerical computation.

A spectrally isolated (possibly degenerate) band is described by a rank-$n$ vector bundle over the Brillouin torus, with fiber $\mathcal{E}(k) \simeq \CC^{n}$ defined by $P(k) \CC^{N}$ where $P(k)$ is the rank-$n$ projector on the degenerate band, and $N$ the total number of bands. 
(For instance, in the situation described in the main text $N=6$ while $n=2$.)
This bundle is equipped with a connection $\nabla^{P} = P \dd$ obtained by projecting the trivial connection on the trivial entire Bloch bundle. Associated to $\nabla^{P}$ is a connection form $A$, explicitly given by equation~\eqref{definition_connection_components} in section~\ref{app_semiclassical}.
The effect of parallel transport in the degenerate band along a path $\mathcal{C}$ on the Brillouin torus is described by the Wilson line operator
\begin{equation}
  W(\mathcal{C}) = \mathcal{P} \exp\left( -\int_{\mathcal{C}} A \right).
\end{equation}
When a basis of each fiber is chosen (such as the generalized polarizations described in section~\ref{app_generalized_polarizations}), the Wilson line operator can be seen as a $n \times n$ matrix.
The curve $\mathcal{C}$ can be seen as a map $[0,1] \to \BZ$ ; it starts at a point $p$ and ends at point $q$ (namely $\mathcal{C}(0) = p$ and $\mathcal{C}(1) = q$).
The Wilson line operator $W(\mathcal{C}) \in \text{End}(\mathcal{E}(p), \mathcal{E}(q))$ is a linear map from the fiber at $p$ to the fiber at $q$, meaning that a state $\phi(p)$ with quasi-momentum $p$ is mapped to a state $\tilde{\phi}(q) = W(\mathcal{C}) \phi(p)$ with quasi-momentum $q$. 
Hence, the Wilson line operator transforms as
\begin{equation}
   W(\mathcal{C}) \mapsto g^{-1}(q) W(\mathcal{C}) g(p).
\end{equation}
under a change of basis $g(k)$ of the fiber at each point. 
When $\mathcal{C}$ is a closed loop (namely $\mathcal{C}(0) = p_0 = \mathcal{C}(1)$), then $W(\mathcal{C})$ transforms covariantly as
\begin{equation}
  W(\mathcal{C}) \mapsto g^{-1}(p_0) W(\mathcal{C}) g(p_0).
\end{equation}
However, this behavior is not independent of the base point $p_0$, i.e. the unitary $g(p_0)$ does depend on $p_0$.

In a numerical computation, $g(p_0)$ is effectively random. While Wilson loops operators can legally be combined if they share an endpoint $k$, the matrices representing them can only be multiplied if they represent operators with the same basis of $\mathcal{E}(k)$. To do so, one must ensure that the same eigenvectors at $k$ are used for both loops, for example by systematically applying a gauge fixing procedure.
We then numerically compute the Wilson line/loop operators using standard techniques \cite{Luscher1982,Panagiotakopoulos1985,Phillips1990,KingSmith1993,Simon1993,Resta1994,Marzari1997,Fukui2005,Leone2011} summarized as follows.
The curve $\mathcal{C}$ is discretized into the discrete path $\hat{\mathcal{C}}$ with vertices
\begin{equation}
   \hat{\mathcal{C}}_{n} \equiv \mathcal{C}\left(\frac{n}{N-1}\right)
\end{equation}
for $n = 0, 1, \dots, N-1$. We then define the overlap matrix
\begin{equation}
  \tilde{S}_{i j}(p,q) = \frac{\braket{\phi_i(p), \phi_j(q)}}{\lVert \phi_i(p) \rVert \, \lVert \phi_j(q) \rVert }.
\end{equation}
In principle, each infinitesimal loop is unitary, but this is only true up to numerical errors. 
Hence, we then use a polar decomposition on $\tilde{S}(p,q) = S(p,q) H_{S}(p,q)$ where $S(p,q)$ is unitary and $H_{S}(p,q)$ is Hermitian positive-definite, and compute
\begin{equation}
  W(\hat{\mathcal{C}}) = \mathcal{P} \prod_{n=1}^{N-1} S(\hat{\mathcal{C}}_{n}, \hat{\mathcal{C}}_{n-1})
\end{equation}
where the $\mathcal{P}$ implies that the product is ordered, so that
$W(\hat{\mathcal{C}}) = S(\hat{\mathcal{C}}_{N-1}, \hat{\mathcal{C}}_{N-2}) \cdots S(\hat{\mathcal{C}}_{1}, \hat{\mathcal{C}}_{0})$.

\section{Generalized polarizations}
\label{app_generalized_polarizations}

Here, we describe a procedure providing a reference for the degenerate generalized polarization states over the Brillouin zone (a choice of gauge).
We consider the situation where the band structure is globally two-fold degenerate, but where the double Dirac cone is gapped, i.e. when at least one mass or family of spring constants is distinct from the others. 
To use the semiclassical equations of motion, one needs to define what the vector $\eta$ describing the composition of a wavepacket means. 
To do so, one has to define a smooth basis of the degenerate subspace at each $k$, at least locally.
There is no obstruction to do so even globally, because the first Chern number of the two-fold degenerate bands vanishes due to time-reversal invariance.
However, a practical way of defining this smooth basis is required. 

\begin{figure}
  \centering
  % generated_by(figure_generalized_polarizations.ipynb)
  \includegraphics{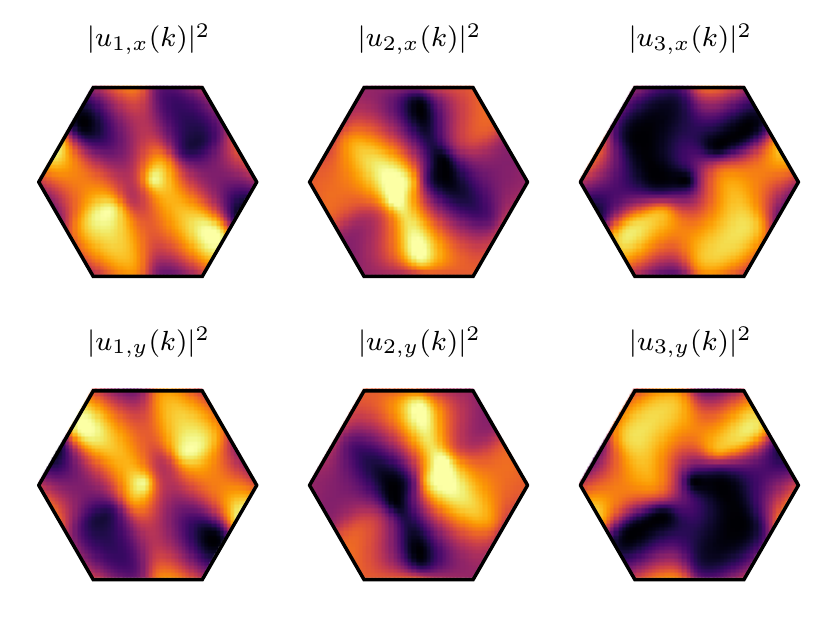}
  \caption{\label{mechanical_spin_up_field}\strong{Generalized polarizations.}
  We plot the amplitude squared of the displacement field $|u^{+}_{M,\mu}(k)|$ for the generalized polarization $+$ over momentum space, where $M=1,2,3$ labels the three masses in the unit cell and $\mu=x,y$  the two components of the displacement.
  The phase of the eigenmodes is not trivial, but is not represented here.
  The corresponding data for the other generalized polarization $-$ is obtained by exchanging $x$ and $y$.
  Here, we have set $\delta m = \num{0.1}$.
  }
\end{figure}

\begin{figure}
  \centering
  \includegraphics{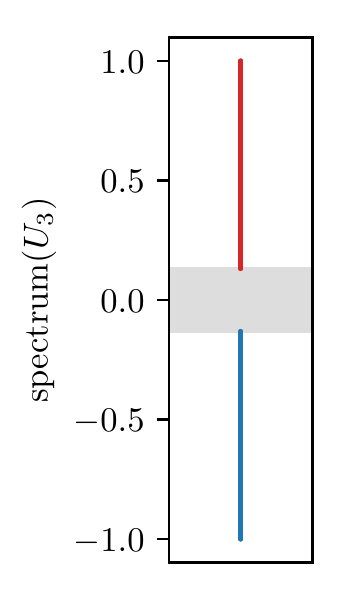}
  \caption{\label{si_figure_distinguishing_generalized_polarizations}\strong{Gauge choice of generalized polarization.}
  The eigenvalues of the projection $U_{3}$ of the self-dual symmetry $(- \ii \mathscr{U}_0)$ on the two-fold degenerate band with dispersion $\omega_3(k) = \omega_4(k)$ are computed on a grid sampling the Brillouin zone.
  A gap (in gray) globally the two eigenvalues (in blue and red).
  This allows to unambiguously define a global frame for the degenerate subspace.
  Here, we have set $\delta m = \num{0.1}$.
  }
\end{figure}

Let $\phi_{i}(k)$ with $i=1,\dots,6$ the six Bloch eigenmodes of the dynamical matrix at momentum $k$, corresponding to the dispersion relations $\omega_{i}(k)$ chosen with $0 < \omega_1(k) = \omega_{2}(k) < \omega_3(k) = \omega_4(k) < \omega_5(k) = \omega_6(k)$. 
Due to the two-fold degeneracy, the eigenstates are not unambiguously defined: any $U(2)$ rotation of the states $(\phi_{i}, \phi_{i+1})$, $i=1,3,5$ is equally acceptable.
In particular, this choice will be random in a numerical diagonalization algorithm.
One can always assume $\braket{\phi_i, \phi_{i+1}} = 0$ ($i=1,3,5$).
If it is not the case, one can orthogonalize the basis of the degenerate eigenspace e.g. through a QR decomposition. 
We can then compute the matrix elements of the duality operator $\mathscr{U}_0$ at $\Gamma$,
\begin{equation}
  U_{i}^{\mu,\nu} = \braket{\phi_{i+\mu}, (- \ii \mathscr{U}_0) \phi_{i+\nu}}
\end{equation}
with $\mu,\nu=0,1$. 
(An additional imaginary factor $\ii$ is used to make the eigenvalues real. Equivalently, we could consider purely imaginary eigenvalues.) 
Focusing on the degenerate band with $i=3$, we observe on figure \ref{si_figure_distinguishing_generalized_polarizations} that a gap separates the two eigenvalues of the $2 \times 2$ reduced matrix $U_{i}$ all over the Brillouin zone.
Hence, defining $\phi_i^{+(-)}(k)$ as the eigenstate of $U_{i}$ with largest (lowest) eigenvalue provides an unambiguous global gauge choice for the degenerate band, that can be seen as a generalized (momentum dependent) polarization.
In figure \ref{mechanical_spin_up_field}, we represent the amplitude of the components of a state in the $+$ polarization over the Brillouin zone.

\section{Orders of magnitude}

\def\Dog{\Delta\omega_{\text{g}}}
\def\Dop{\Delta\omega_{\text{p}}}

In this section, we discuss some of the approximations used in the semi-classical analysis of section~\ref{app_mechanical_bloch_oscillations}.
We take into account the following:
\begin{itemize}
  \item the wave packet has to both have well-defined semi-classical position and momentum
  \item the duration of the experiment is set by the need of traversing the Brillouin zone under a constant force;
  \item the force has to be small enough so that non-adiabatic Landau-Zener transitions can be neglected;
  \item the breaking of the two-fold degeneracy has to be small enough so that the bands are mixed;
  \item the dissipation has to be small enough so that the wave packet is still measurable after the process.
\end{itemize}

First, the size of the wave packet in momentum space $w_k$ should be small with respect to the size of the Brillouin zone $2\pi/a$ (and with respect to the variations inside the Brillouin zone), and large with respect to the discrete grid $2\pi/L$ due to the finite size $L$ of the system in real space, so that the wave packet also has a well-defined semi-classical position. Hence, one should have
\begin{equation}
  \frac{2\pi}{L} \ll w_k \ll \frac{2\pi}{a}.
\end{equation}

Integrating the equation $\dot{k} = - \nabla V$ for a uniform force $F = - \nabla V$, we obtain $k = k_0 + F t$, where $k_0$ is the initial momentum and $t$ the time elapsed since the beginning of the process. The duration $T$ of the process is chosen so that $F T$ is a reciprocal lattice vector, giving $F T = 2 \pi / a$ where $a$ is the lattice spacing. This gives a order of magnitude of the duration of the experiment $T = \frac{2 \pi}{F a}$.
It is convenient to write $F = \Dop / a$, where $\Dop$ is the characteristic change in frequency from a unit cell to the neighboring one due to the external potential, leading to
\begin{equation}
  T = \frac{2 \pi}{\Dop}.
\end{equation}

Now, one can estimate the non-adiabatic tunneling probability $P$ from the Landau-Zener formula \cite{Landau1932,Zener1932,Stueckelberg1932,Majorana1932} (see also \cite{Takahashi2017} and \cite{Pruneda2009} for Bloch states) $P = \exp\left(-2 \pi \Gamma\right)$ where $\Gamma = \frac{\Delta^2}{2 F v}$ for a one-dimensional Dirac Hamiltonian of the form $H(k) = v k \sigma_z + \Delta \sigma_x$, where $v$ is a characteristic group velocity and $\Delta$ is the gap. As we wish to neglect non-adiabatic transitions, $\Gamma$ has to be as big as possible.

To estimate $v$ and $\Delta$, we use the effective dynamical matrix in section \ref{app_effective_dynamical_matrix} and expand the square root of the spectrum $[(\omega/\omega_0)^2](k) = 3 + f(k)$ at first order to find $v \simeq \omega_0/8 \times a/2\pi$ and $\Delta \simeq M \omega_0/8$ 
Here, $\omega_0$ is a dimensionful characteristic angular frequency, $M$ is dimensionless, and $a$ is the length of a lattice vector. 
(In section \ref{app_effective_dynamical_matrix}, the effective dynamical matrix and the wave vector $k$ are dimensionless.)
Using $F = \Dop / a$, we are led to
\begin{equation}
\Gamma \equiv \frac{2 \pi}{16} \, \frac{M^2}{(\Dop / \omega_0)} > 1
\end{equation}
where the inequality represents our wish of avoiding non-adiabatic transitions.

Note that the group velocity also gives an idea of the distance traveled by a wave packet $d = v T \simeq (\omega_0/\Dop) a/8$ (this is indeed complicated by the anomalous velocity, but gives an approximate order of magnitude), and the system size has to be at least as big as $d$, so we also have
\begin{equation}
  \frac{L}{a} > \frac{\omega_0}{8 \Dop}.
\end{equation}

Additional constraints stem from likely experimental constraints. 
One can expect that an experimental realization will suffer from a small breaking of the duality, e.g. due to effective springs connecting second-nearest neighbors. 
This would for instance be the case in the 3D printed system of reference \cite{Ma2018}.
Similarly, the value of the twisting angle might be only approximately equal to the critical one.
Hence, the two-fold degeneracy will be weakly lifted, and we write $\Dog$ the order of magnitude of the lifting.
In order to be able to consider that the bands are effectively degenerate, the lifting should be small enough so that non-adiabatic transitions do occur between them (see \cite{Segert1987,Comparat2009}). Using the Landau-Zener formula with $\Dog$ as the gap, we require
\begin{equation}
  \Gamma' \equiv 8 \pi \, \frac{\Dog^2}{\Dop \omega_0} \ll 1.
\end{equation}
Additionally, we expect that phase coherence between the almost degenerate modes should be preserved (so they can effectively be considered to have the same dispersion relation), requiring $\Dog T \ll 2 \pi$ where $T$ is the duration of the experiment.

Besides, one can expect that various dissipative processes will effectively produce a uniform damping rate $\gamma$ (appearing in the equations of motion as $\partial_t^2 \ket{\phi} = \hat{D} \ket{\phi} - \gamma \Id \partial_t \ket{\phi})$, leading to the attenuation of the wave packet.
In order to still be able to observe the wave packet at the end of the experiment, one requires
\begin{equation}
  \gamma T \lesssim 1.
\end{equation}
Indeed, the precise constraint would depend on the efficiency of the measurement system.

\section{Relation between the first and second order formalisms}
\label{app_relation_first_second_order}

In this section, we discuss the relation between the first-order and the second-order formalisms,
and relate the Berry connections obtained from the dynamical matrix and from the first-order Hamiltonian-like operator.
Our simplified discussion relies on strong hypotheses, relevant in the case at hand but generally not satisfied (in particular, we assume no dissipation; or in a slightly less restrictive way that the dissipation is trivial, in the sense that it is represented by a scalar matrix).
A more general discussion on Berry phases for non-Hermitian Hamiltonians can be found in references \cite{Miniatura1990,
Dattoli1990} and for a situation closer to the present one in references \cite{Wang2009,Zhang2010}.
A different but equivalent point of view is used in references \cite{Onoda2006,Zhang2010b,Inoue2019} where the semi-classical equations of motion are also discussed.

The linear(ized) Newton equations of motion
\begin{equation}
  \partial_t^2 \phi = D \phi 
\end{equation}
are equivalent to the linear(ized) Hamilton canonical equations of motion
\begin{equation}
  \label{first_order_eom}
  \partial_t \psi = L \psi
\end{equation}
where
\begin{equation}
  L = \begin{pmatrix} 0 & 1 \\ - D & 0 \end{pmatrix}
\end{equation}
and where $\psi=(\phi,\dot{\phi})$.
The canonical Hamilton equations are arguably more fundamental.
Most importantly, the semi-classical analysis described in section~\ref{app_semiclassical} relies on the first-order equation \eqref{first_order_eom}.

Here, the dynamical matrix $D = D^{\dagger}$ is assumed to be Hermitian.
(This is the case for the mechanical system we study in the main text.)
Indeed, the operator $L$ is not Hermitian (neither is $H = \ii L$).
However, the eigenvectors and eigenvalues of $D$ and $L$ (or $H$) are in one-to-two correspondence except when the eigenvalues vanish. 
Let us show that directly. 
For simplicity, we assume that $D$ is positive definite.
As $D$ is Hermitian, it is diagonalizable, so let $(\omega_i^2, \phi_i)$ be a basis of orthonormal eigenvectors
such that $D \phi_i = \omega_i^2 \phi_i$, where $\omega_i > 0$.

As $L$ (and $H=\ii L$) are not Hermitian, we need to consider a biorthogonal system of right and left eigenvectors instead of a orthonormal family of eigenvectors.
Hence, let us define
\begin{equation}
  \ket{\psi_i^{\pm}} = \begin{pmatrix}
   \phi_i \\ 
   \pm \ii \omega_i \phi_i 
   \end{pmatrix}
\end{equation}
and
\begin{equation}
  \bra{\tilde{\psi}_i^{\pm}} = \frac{1}{\pm 2 \ii \omega_i} (\pm \ii \omega_i \phi_i^\dagger, \phi_i^\dagger).
\end{equation}
so that
\begin{subequations}
\begin{align}
  &L \ket{\psi_i^{\pm}} = (\pm \ii \omega_i) \ket{\psi_i^{\pm}} \\
  \bra{\tilde\psi_i^{\pm}} &L = \bra{\tilde\psi_i^{\pm}} (\pm \ii \omega_i).
\end{align}
\end{subequations}
For $H = \ii L$, $\psi_{\pm}$ corresponds to the real eigenvalues $\mp \omega$.
Note that the definition of $\psi_i^{\pm}$ and $\tilde{\psi}_i^{\pm}$ is not unique, as we could e.g. make the normalization more symmetric.
The states $\psi_i^{\pm}$ are linearly independent, except for zero-frequency mode of the dynamical matrix with $\omega_i^2 = 0$, where they are equal. This case corresponds to an exceptional point \cite{Kato1984} where $L$ is not diagonalizable.
Here, we focus on finite frequency modes, for which the states $\psi_i^{\pm}$ are linearly independent, and satisfy
\begin{equation}
  \braket{\tilde{\psi}_{i}^{\pm}, \psi_{j}^{\pm}} = \delta_{i, j}
  \quad \text{and} \quad
  \braket{\tilde{\psi}_{i}^{\pm}, \psi_{j}^{\mp}} = 0.
\end{equation}

\subsection{Berry connection}

Let us now focus on a subset of degenerate states (sharing the same $\omega = \omega_{i}$ for all $i$ in the family) and let us compute the Berry connection
\begin{equation}
  \mathcal{A}_{i,j}^{\pm,\pm}
  = \braket{\tilde\psi_{i}^{\pm}, \dd \psi_{j}^{\pm}} 
  = \braket{\phi_{i}, \dd \phi_{j}} 
  + \frac{\delta_{i,j}}{2} \dd \log \omega.
\end{equation}
The first term $A_{i,j} = \braket{\phi_{i}, \dd \phi_{j}}$ is the Berry connection of the eigenvectors of the dynamical matrix.
The second term is a total derivative (here it is crucial that $\omega$ remains strictly positive), and can be ignored:
it can be absorbed by an appropriate gauge choice, and does not modify gauge-invariant quantities.
Similarly, we can compute the cross terms between positive and negative frequencies,
\begin{equation}
  \mathcal{A}_{i,j}^{\pm,\mp}
  = \braket{\tilde\psi_{i}^{\pm}, \dd \psi_{j}^{\mp}} 
  = \frac{\delta_{i,j}}{2} \dd \log \omega
\end{equation}
that are pure gauge, and therefore ignored.
Hence, we can focus on the Berry connection $A_{i,j}$ obtained from the dynamical matrix.

\subsection{Dualities}

When there is no dissipation, the duality on the dynamical matrix directly translates into a duality for the first-order (Hamiltonian like) operators.
Assume that $\hat{\mathscr{U}} \hat{D} \hat{\mathscr{U}}^{-1} = \hat{D}^*$.
Let $\hat{L}$ and $\hat{L}^*$ be the first-order operators corresponding to $\hat{D}$ and $\hat{D}^*$, respectively.
Finally, let
\begin{equation}
  \hat{\mathcal{U}} = \begin{pmatrix}
    \hat{\mathscr{U}} & 0 \\
    0 & \hat{\mathscr{U}}
  \end{pmatrix}.
\end{equation}
Then, we have $\hat{\mathcal{U}} \, \hat{L} \, \hat{\mathcal{U}}^{-1} = \hat{L}^*$.

The duality can also be extended to the cases where the dissipation term is trivial (so it commutes with $\hat{\mathscr{U}}$), or when there is also a duality for the dissipation operator.

\section{Comparison with a non-symmorphic symmetry}

\begin{figure*}
  \centering
  \includegraphics{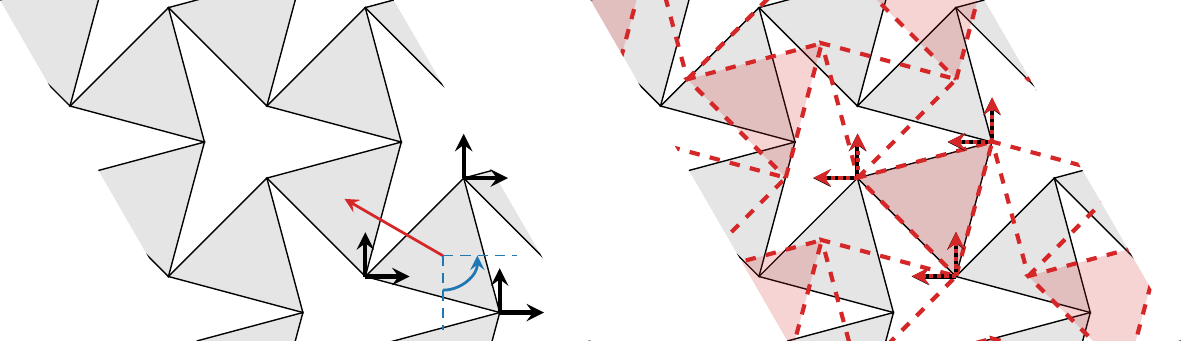}
  \caption{\label{si_figure_effect_nonsymmorphic}\strong{Effect of a quarter rotation followed by a non-integer translation.}
  }
\end{figure*}

At first sight, the self-dual symmetry resembles the non-symmorphic operation $\{\ang{90} | (-1/3, 1/3) \}$.
However, the two operations are different. 
This is expected, because the space group of the twisted Kagome lattice does not contain the spatial transformation $\{\ang{90} | (-1/3, 1/3) \}$.
On figure \ref{si_figure_effect_nonsymmorphic}, we superimpose the original lattice with the transformed lattice; it is visible that they differ, except on a single triangle.

\section{Effective dynamical matrix}
\label{app_effective_dynamical_matrix}

A $k\!\cdot\!p$-like expansion can be used to understand the effect of different values for the masses in the unit cell.
The matrix $D(\Gamma)$ has dimensionless eigenvalues $(0,0,3,3,3,3)$. 
The four degenerate eigenvalues correspond to the double Dirac cone; let $\phi_i$ with $i=3,\dots,6$ the corresponding eigenvectors, and compute the $4 \times 4$ effective dynamical matrix
\begin{equation}
  [D_{\text{eff}}(k)]_{i j} = \braket{\phi_i, D(k, \delta \mu) \phi_j}
\end{equation}
with $i,j=3,\dots,6$,
and where $D(k, \delta \mu)$ is the dynamical matrix where the mass $m_1$ is perturbed so that $\mu_1 \equiv \sqrt{m_1} = 1 + \delta \mu$, and all others masses are set to unity.
We find
\begin{equation}
\begin{split}
  D_{\text{eff}}(k ; \delta \mu) = D_0 + D_1(k) + D_2(k) + \mathcal{O}(k^3) \\
  + \delta\mu \, \left[ E_0 + E_1(k) + E_2(k) + \mathcal{O}(k^3) \right] + \mathcal{O}(\delta \mu^3)
\end{split}
\end{equation}
where $D_0$ and $E_0$ do not depend on $k$, $D_1(k)$ and $E_1(k)$ are linear in $k$, $D_2(k)$ and $E_2(k)$ are (homogeneous) quadratic in $k$, etc. 
After an appropriate unitary rotation of the basis,
\begin{equation}
\begin{split}
  D_{\text{eff}}(k ; \delta \mu) = (3 + \frac{\sqrt{3}}{4} M) \Id_{4} \\
  - \frac{\sqrt{3}}{4} 
  \begin{pmatrix}
    k \cdot \sigma + M \sigma_{z} & M \Delta(k) \\
    M \Delta^{\dagger}(k) & k \cdot \sigma - M \sigma_{z}
  \end{pmatrix} + \mathcal{O}(k^2),
\end{split}
\end{equation}
where $k \cdot \sigma = k_x \sigma_x + k_y \sigma_y$ and
\begin{equation}
\begin{split}
  \Delta(k) = \frac{\ii}{48} \big( 2 (\sqrt{3} k_x + 3 k_y) \sigma_0 \\
   + (-5 \sqrt{3} k_x + 3 k_y) \sigma_x 
   + 3 (- k_x + \sqrt{3} k_y) \sigma_z \big)
\end{split}
\end{equation}
where we set $M = - 8/\sqrt{3} \delta \mu$. 

For a general perturbation of the masses $\mu_i \equiv \sqrt{m_i} = 1 + \delta_i$, the term $E_0$ becomes
\begin{equation*}
\footnotesize
\begin{pmatrix}
   -3 (\delta_2+\delta_3) & 0 & \sqrt{3} (\delta_2-\delta_3) & 0 \\
 0 & -4 \delta_1-\delta_2-\delta_3 & 0 & \sqrt{3} (\delta_3-\delta_2) \\
 \sqrt{3} (\delta_2-\delta_3) & 0 & -4 \delta_1-\delta_2-\delta_3 & 0 \\
 0 & \sqrt{3} (\delta_3-\delta_2) & 0 & -3 (\delta_2+\delta_3) \\
 \end{pmatrix}
\end{equation*}

\section{Semi-classical dynamics}
\label{app_semiclassical}

In the context of wave physics, semi-classical approximations provide an approximate particle-like description of a wave packet localized both in physical space and momentum space. For instance, geometrical optics can be viewed as a short-wavelength approximation of Maxwell equations~\cite{Born1999}. 
The equations describing the semi-classical dynamics of a wave packet in a (perturbed) spatially periodic structure can be systematically obtained from the underlying wave equations~\cite{Karplus1954,Chang1996,Sundaram1999,Panati2003,Shindou2005,Culcer2005,Chang2008,Xiao2010}. This method was first and foremost applied to electrons in solids, but applies to all waves. It was used to describe the semi-classical dynamics of light waves \cite{Onoda2004,Bliokh2004,Onoda2006,Bliokh2007} and to transverse acoustic waves \cite{Bliokh2006,Mehrafarin2009,Torabi2009}. 

Here, we consider a wave packet constrained to evolve in a set of degenerate bands in a critical Kagome lattice. An acoustic wavepacket centered at the semi-classical position $r$ and momentum $k$ is defined as the superposition
\begin{equation}
  \label{definition_wavepacket}
  \ket{\Psi(r,k)} = \sum_{i} \frac{1}{\lVert \BZ \rVert} \int_{\BZ} \dd^2 q \; a(t, q) \eta_{i}(t, q) \ket{\phi_{i}(q)}
\end{equation}
of Bloch states $\ket{\phi_{i}(k)}$ where $i = i_1, \dots, i_g$ label the $g$ bands involved in the wavepacket (in the system discussed in the main text, $g=2$; for simplicity, we will write $i = 1, \dots, g$ in the following).
Here, $\eta=(\eta_1, \dots, \eta_g)^{T}$ is a normalized vector describing the band composition of the wave packet,
while $a$ is a narrow distribution centered at the semi-classical momentum $k$ (while its Fourier transform is centered at the semi-classical position $r$), normalized as
\begin{equation}
  \frac{1}{\lVert \BZ \rVert} \int_{\BZ} \dd^2 q \; a(t, q) = 1
\end{equation}
so that $\braket{\Psi\mid \Psi} = 1$. 
When there are degeneracies, the assignment of the band indices $i$ is not trivial. For concreteness, let us focus on the case of two globally degenerate bands. One has to define a smooth frame of the degenerate vector bundle, i.e. one has to decide which band is called $\ket{\phi_{1}(k)}$ and which band is called $\ket{\phi_{2}(k)}$ at each momentum $k$, in a continuous fashion. 
In general, it may only possible to so locally, e.g. when a band carries a Chern number.
In the situation discussed in the main text, (momentum-dependent) generalized mechanical polarizations $\pm$ can be globally defined as described in section~\ref{app_generalized_polarizations}.
Note that the generalized polarizations might not have a specific physical meaning.
In particular, there is still mode interconversion between the modes $\pm$ under an external force (this is the case in any basis due to the non-Abelian nature of the Berry connection).

The semiclassical evolution of the wavepacket is described by the semiclassical Lagrangian \cite{Culcer2005,Onoda2006,Bliokh2007,Bliokh2006}
\begin{equation}
  \mathcal{L}(r,k,\eta,\dot{r},\dot{k},\dot{\eta}) = \braket{\Psi \mid \left( \ii \partial_t - \ii L \right) \mid \Psi}
\end{equation}
obtained from the first-order equations of motion \eqref{first_order_eom} applied to the semiclassical wavepacket \eqref{definition_wavepacket} described by the semiclassical variables $(r,k,\eta,\dot{r})$ and their derivatives.
Here, $\ii L = H$ is the linear operator describing the first-order dynamics of the mechanical system, equivalent to a Hamiltonian.
After simplification, the semiclassical Lagrangian reads \cite{Culcer2005,Onoda2006,Bliokh2007,Bliokh2006}
% ATTENTION AUX \ii, mon A_ij = <psi_i,d psi_j> et a une courbure F = dA + A \wedge A
\begin{equation}
  \mathcal{L} = \braket{\eta, ( \ii \partial_t + \ii A(k) \cdot \dot{k} + k \cdot \dot{r} - \mathcal{H}(r,k) ) \, \eta}. 
  % = \mathcal{L}(r,k,\eta,\dot{r},\dot{k},\dot{\eta})
\end{equation}
In this effective Lagrangian, $r$ and $k$ are the position and momentum of the center of mass of the wave packet, $\eta=(\eta_1, \dots, \eta_g)^{T}$ is a normalized vector describing the band composition of the wave packet, $\mathcal{H}(r,k) = \Omega(k) + V(r, k)$ is a semiclassical Hamiltonian composed of the degenerate bulk band dispersion relation $\Omega(k)$ and of an external potential $V(r,k)$. 
Finally, $A(k)$ and $F(k)$ are the matrix-valued non-Abelian Berry connection and curvature forms of the degenerate band, respectively defined as
\begin{equation}
  A = A^{\mu} \dd k_{\mu}
\end{equation}
where $A^{\mu}$ is an operator with matrix elements
\begin{equation}
  \label{definition_connection_components}
  A^{\mu}_{i j}  = \braket{\phi_i \mid \partial^{\mu} \phi_j}
\end{equation}
and
\begin{equation}
  F = \dd A + A \wedge A = F^{\mu \nu} \dd k_{\mu}
\end{equation}
where $F^{\mu \nu}$ is an operator with matrix elements
\begin{equation}
  \label{definition_curvature_components}
  F_{i j}^{\mu \nu} = \partial A_{i j}^{\nu} / \partial k_{\mu} - \partial A_{i j}^{\mu} / \partial k_{\nu} + A_{i k}^{\mu} A_{k j}^{\nu} - A_{i k}^{\nu} A_{k j}^{\mu}.
\end{equation}

The corresponding semi-classical equations of motion then read
\begin{equation}
\label{semiclassical_eom}
\begin{split}
  \dot{r}^{\mu} &=  \Braket{\eta, \left( \left[ \frac{\partial}{\partial k_{\mu}} - A^{\mu}, \mathcal{H} \right] + \ii F^{\mu \nu} \, \dot{k}_{\nu} \right) \, \eta } \\
  \dot{k}_{\mu} &= - \Braket{\eta, \frac{\partial \mathcal{H}}{\partial r^{\mu}} \eta} \\
  \dot{\eta}    &= - (\ii \mathcal{H} + A^{\mu} \dot{k}_{\mu}) \, \eta.
\end{split}
\end{equation}
This description assumes a single well-defined wavepacket (e.g. it does not describe the splitting of a wavepacket into several parts), and it ignores non-adiabatic Landau-Zener transitions \cite{Landau1932,Zener1932,Stueckelberg1932,Majorana1932} (see \cite{Takahashi2017} and \cite{Pruneda2009} for a discussion in the context of Bloch states).

Non-adiabatic transitions are negligible as long as the external force $f = - \partial_r \mathcal{H}$ is small enough, mainly compared to the gap (see references for details).
When such transitions occur, the bands are mixed; in this situation, one can study Landau-Zener-Stückelberg interferences \cite{Lim2012,Lim2015,Li2016}. 
The structure of our system suggests that the LZS interferences may also be non-Abelian, but their analysis is outside of the scope of this work.

In general, the three equations of motion are coupled. Note however that if we are given the trajectory $\mathcal{C}$ of the wavepacket in momentum space, the change of composition is obtained as the path-ordered exponential
\begin{equation}
  \eta(t) = P \exp\left( - \int_{\mathcal{C}} (\ii \mathcal{H} + A^{\mu} \dot{k}_{\mu}) \dd k \right) \eta(0).
\end{equation}
Assuming that the quantity $\mathcal{H}$ is scalar, this is split into a dynamical phase $\ee^{-\ii \mathcal{H} t}$ (with a time-independent $\mathcal{H}$), and a geometric phase
\begin{equation}
  W[\mathcal{C}] = P \exp\left( - \int_{\mathcal{C}} A \right).
\end{equation}
called a Wilson line operator, that is a property of the non-Abelian Berry connection. 
Further details on Wilson lines and Wilson loops are discussed in section~\ref{app_wilson_loops}.

\section{Deformed Kagome lattices}
\label{app_deformed_kagome}

In this paragraph, we recall the definition of twisted and deformed Kagome lattices, and specify the conventions we use throughout the paper.
We also explore the relation between the self-duality and the existence of orthogonal bonds in the mechanical network.
Finally, we analyze the space group symmetries of the Kagome lattices; in particular, we find that the wallpaper group (2D space group) of twisted Kagome lattices is \hm{p31m} as long as the twisting angle is nonzero, and does not change at the critical twisting angle. This invalidates any potential explanation of the particular properties of the critical lattice in terms of spatial symmetries.

\subsection{General description}

\begin{figure}
  \centering
  \includegraphics[width=11cm]{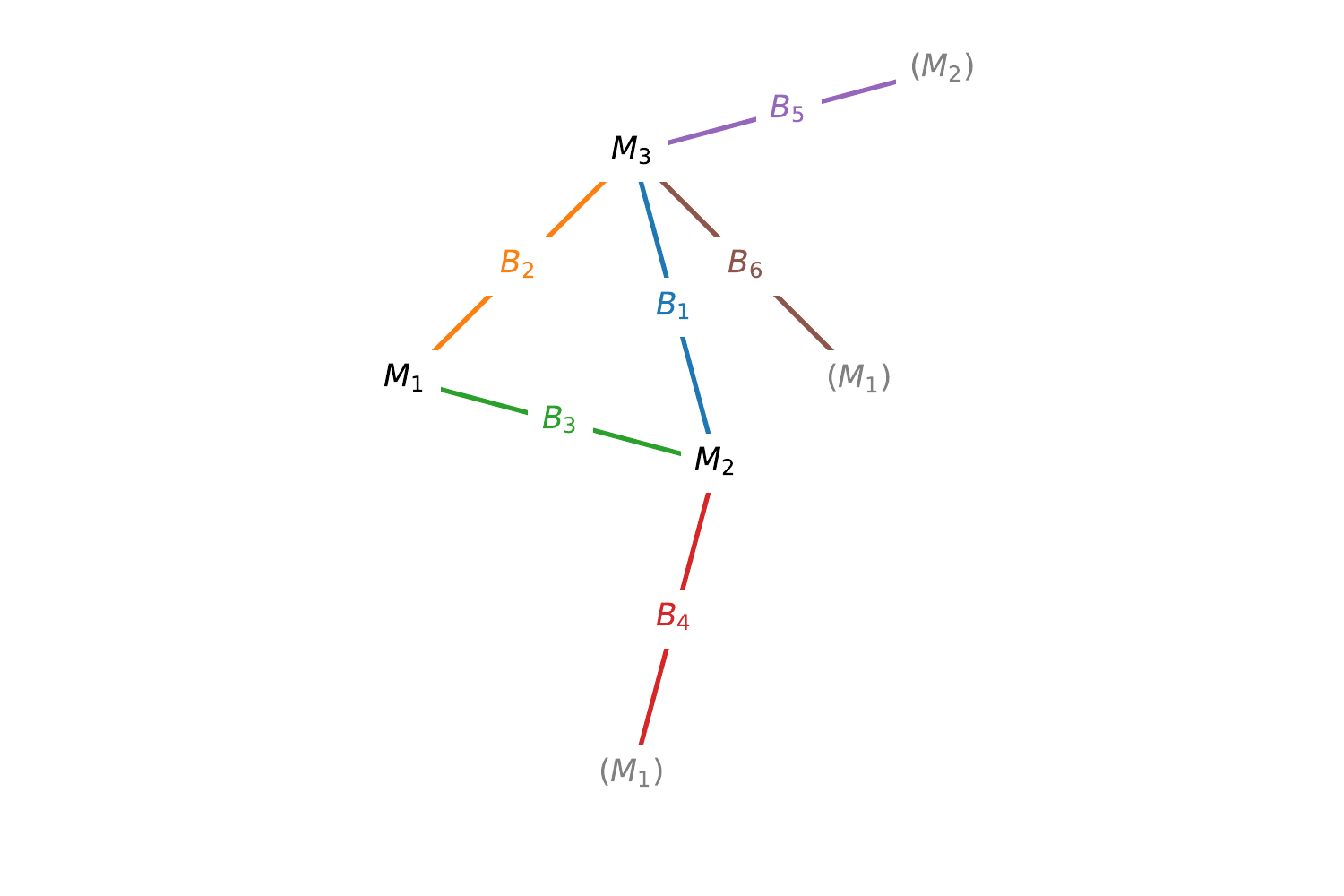}
  \caption{\label{unit_cell_kagome}\strong{Unit cell of the Kagome lattices.} }
\end{figure}

In the main text, we consider mechanical Kagome lattices \cite{Hyun2002,Guest2003,Hutchinson2003,Hutchinson2006,Souslov2009,Sun2012,Kane2013,Lubensky2015}, two-dimensional mechanical lattices composed of three masses per unit cell (see figure \ref{unit_cell_kagome}), connected by harmonic springs. 
They were notably used as an example of collapse mechanism \cite{Guest2003,Hutchinson2003} and to demonstrate the existence of topological properties at zero frequency in mechanical band structures \cite{Kane2013}. 
Kagome lattices are isostatic, meaning that there are as many degrees of freedom ($3 \times 2 = 6$ possible displacements per unit cell) than there are constraints ($6$ springs per unit cell). 

Following references~\cite{Souslov2009,Sun2012,Kane2013}, we consider \emph{deformed Kagome lattices}, defined on a triangular Bravais lattice with primitive vectors chosen as
\begin{subequations}
\begin{align}
  a_1 &= [1, 0]^T \\
  a_2 &= [\cos(2\pi/3), \sin(2\pi/3)]^T = [-1/2,\sqrt{3}/2]^T.
\end{align}
\end{subequations}
To describe a physical system, both primitive vectors should be multiplied with a length $a$, which here is taken to the unity for simplicity. We also define $a_3 = - a_1 - a_2$ for convenience. Besides, it will be useful to write $\kappa_i = k \cdot a_i$ where $k=(k_x, k_y)$ is a reciprocal vector.

Deformed Kagome lattices are a family of elastic networks described by four real parameters $(x_1,x_2,x_3,z)$ as follows. First, define
\begin{equation}
\begin{split}
    y_1 &= \frac{z}{3} + x_3 - x_2 \\
    y_2 &= \frac{z}{3} + x_1 - x_3 \\
    y_3 &= \frac{z}{3} + x_2 - x_1
\end{split}
\end{equation}
and
\begin{equation}
\begin{split}
  s_1 &= x_1 (a_3-a_2)+y_1 a_1 \\
  s_2 &= x_2 (a_1-a_3)+y_2 a_2 \\
  s_3 &= x_3 (a_2-a_1)+y_3 a_3.
\end{split}
\end{equation}

The unit cell of the deformed Kagome lattice is composed of three masses $M_i$, $i=1,2,3$. Their positions can be chosen as 
\begin{equation}
\begin{split}
  r_1 &= -a_3/2 \\
  r_2 &= a_1/2 + s_2 \\
  r_3 &= (a_1-a_3)/2 - s_1.
\end{split}
\end{equation}
Each mass has a physical mass $m_i$ (which we consider to be dimensionless), which is taken to be the unity $m_i=1$ unless otherwise specified.

The masses are connected by bonds $B_a$. In a spatially periodic system, we also have to choose a unit cell for the bonds, which are described by the data $(M_i, M_j, \gamma_{i j})$ of the two families of masses $M_i$ and $M_j$ connected by the bond, and a Bravais lattice vector $\gamma_{i j} \in \Gamma$ describing the distance between the masses in the lattice. For convenience, we also write $\text{start}(B_a) = M_i$, $\text{end}(B_a) = M_k$ and $\text{jump}(B_a) = \gamma_{i j}$.
Here, we choose
\begin{equation}
\begin{split}
  B_1 &= (M_2, M_3, 0), \\
  B_2 &= (M_3, M_1, 0), \\
  B_3 &= (M_1, M_2, 0), \\
  B_4 &= (M_2, M_1, -a_2), \\
  B_5 &= (M_3, M_2, a_1+a_2), \\
  B_6 &= (M_3, M_1, a_1).
\end{split}
\end{equation}
Additionally, each bond $B_a$ has a spring stiffness $k_a$ which is taken to be the unity $k_a = 1$ unless otherwise specified. Here, we assume that the bonds are at their rest length when the system is at mechanical equilibrium.

\subsection{Twisted Kagome lattices}

Twisted Kagome lattices are a one parameter sub-family of deformed Kagome lattices with
\begin{equation}
  (x_1,x_2,x_3,z)=(x,x,x,0)
\end{equation}
where $x$ is called the twisting parameter. 
They can be equivalently described by the twisting angle
\begin{equation}
  \theta = \arctan(2 \sqrt{3} x).
\end{equation}

\medskip

In this discussion, we have assumed the primitive lattice vectors to have length unity. 
Starting from a undeformed Kagome lattice [with $(x_1,x_2,x_3,z)=(0,0,0,0)$], one can physically access twisted Kagome lattices through the Guest-Hutchinson mechanism.

It should be noted that the lattice constant changes from $a_0$ to $a(\theta) = a_0 \cos(\theta)$ when doing so~\cite{Souslov2009,Sun2012,Kane2013}. 
We follow the convention from \cite{Souslov2009,Sun2012,Kane2013} where a unit lattice constant is assumed.
In a physical implementation of the mechanical system, this assumption is only valid at one point at a time in parameter space.
In particular, this leads to a global rescaling of the frequency spectrum (i.e. the characteristic angular frequency $\omega_0$ used to nondimensionalize the frequency spectrum would depend on $\theta$ through $a(\theta)$).
In this case, it might be convenient to write the duality relation as $\hat{\mathscr{U}}(\theta) \tilde{D}(\theta^*) \hat{\mathscr{U}}(\theta)^{-1} = f(\theta) \tilde{D}(\theta)$ where $\tilde{D}(\theta^*)$ is the dimensionful dynamical matrix, and where $f(\theta) = \omega_0(\theta^*)/\omega_0(\theta)$ is a scalar factor.

\subsection{Critical lattices}
\label{app_critical_lattices}

We observed that the critical twisted Kagome lattice with $\theta = \theta_{\text{c}}$ has three pairs of orthogonal bonds.
This geometric property appears to be intimately related to the self-duality. 
There is a particular sub-family of deformed Kagome lattices (including the critical twisted Kagome lattice as a particular case), that share both this property, and are all self-dual in the sense of equation \eqref{unitary_duality}.
Geometrically, the existence of three pairs of orthogonal bonds occurs when the parameters $(x_1,x_2,x_3,z)$ are roots of the set of three second order polynomials
% generated_by(kagome_right_angles_conditions.ipynb)
\begin{align*}
  %0 &= x_{1}^{2} - 2 x_{1} x_{2} - \frac{2 x_{1}}{3} z + x_{2}^{2} + \frac{2 x_{2}}{3} z + 3 x_{3}^{2} + \frac{z^{2}}{9} - \frac{1}{4} \\
  %0 &= 3 x_{1}^{2} - \frac{1}{36} \left(- 6 x_{2} + 6 x_{3} + 2 z + 3\right) \left(6 x_{2} - 6 x_{3} - 2 z + 3\right) \\
  %0 &= x_{1}^{2} - 2 x_{1} x_{3} + \frac{2 x_{1}}{3} z + 3 x_{2}^{2} + x_{3}^{2} - \frac{2 x_{3}}{3} z + \frac{z^{2}}{9} - \frac{1}{4}.
  - x_{1}^{2} + 2 x_{1} x_{2} + \frac{2 x_{1}}{3} z - x_{2}^{2} - \frac{2 x_{2}}{3} z - 3 x_{3}^{2} - \frac{z^{2}}{9} + \frac{1}{4}\\
  3 x_{1}^{2} + x_{2}^{2} - 2 x_{2} x_{3} - \frac{2 x_{2}}{3} z + x_{3}^{2} + \frac{2 x_{3}}{3} z + \frac{z^{2}}{9} - \frac{1}{4}\\
  - x_{1}^{2} + 2 x_{1} x_{3} - \frac{2 x_{1}}{3} z - 3 x_{2}^{2} - x_{3}^{2} + \frac{2 x_{3}}{3} z - \frac{z^{2}}{9} + \frac{1}{4}
\end{align*}
found by defining $\delta(B) = \text{start}(B) - (\text{end}(B) + \text{jump(B)})$ and computing the scalar products $\braket{\delta(B), \delta(B')}$ for all couples $(B,B')$ of bonds.
If one excludes the degenerate cases where two distinct bonds collapse together, the solution of this system is located in the plane $(x,x,x,z)$ [so $x_1 = x_2 = x_3$], and is determined by the equation
\begin{equation}
  3 x^2 + \frac{z^2}{9} - \frac{1}{4} = 0.
\end{equation}
The solutions can be expressed as two branches $(x, x, x, \pm z_{\text{c}}(x))$, where
\begin{equation}
  z_{\text{c}}(x) = 3 \sqrt{3} \sqrt{x_{\text{c}}^2 - x^2}
\end{equation}
and where $x_{\text{c}} = \sqrt{3}/6$ corresponds to the critical twisting angle $\theta_{\text{c}} = \pi/4$. 
Hence, the critical lattices form an ellipse in the parameter space $(x_1,x_2,x_3,z)$ of deformed Kagome lattices.

Note that the self-duality is a geometric feature, in that it is preserved when the physical masses $m_i$ are changed. It is also preserved when the spring constants are modified, provided that the unit cell is not enlarged (the bonds connecting the same families of masses must have the same spring constants, namely $k_1 = k_5$, $k_2 = k_6$, and $k_3 = k_4$).

\subsection{Determination of the plane group symmetry of deformed Kagome lattices}
\label{app_sec_kagome_plane_groups}

\begin{table}
\centering 
\begin{tabular}{ccccc} 
  \toprule
  Kagome lattice & standard & twisted & deformed & deformed \\
  parameters & $(0,0,0;0)$ & $(x,x,x;0)$ & $(x,x,x;z)$ & $(x_1,x_2,x_3;z)$ \\
  \midrule%
  $M_1 \sim M_2 \sim M_3$ & \hm{p6mm} & \hm{p31m} & \hm{p3} & \hm{p1} \\
  $M_1 \sim M_2 \neq M_3$ & \hm{c2mm} & \hm{c1m1} & \hm{p1} & \hm{p1} \\
  $M_1 \neq M_2 \neq M_3$ & \hm{p2} & \hm{p1} & \hm{p1} & \hm{p1} \\
  \bottomrule%
\end{tabular}
\caption{\label{plane_groups_structures}\strong{Plane groups of the deformed Kagome lattices.}
The label $M_1 \neq M_2 \neq M_3$ is meant to imply that all masses are inequivalent, so we also have $M_1 \neq M_3$.
}
\end{table}

\begin{table}
\centering 
\begin{tabular}{ccc} 
  \toprule
  plane group & point group at $\Gamma$ & Sch. \\
  \midrule
  \hm{p6mm} & \hm{6mm} & \sch{C6v} \\ % 3m at K
  \hm{p31m} & \hm{3m} & \sch{C3v} \\ % 3m at K
  \hm{p3} & \hm{3} & \sch{C3} \\
  \hm{p1} & \hm{1} & \sch{C1} \\
  \hm{c2mm} & \hm{2mm} & \sch{C2v} \\
  \hm{c1m1} & \hm{m} & \sch{Cs} \\
  \hm{p2} & \hm{2} & \sch{C2} \\
  \bottomrule%
\end{tabular}
\caption{\label{points_groups_at_Gamma}\strong{Points groups at the Gamma point.}
}
\end{table}

\begin{figure}
  \centering
  \includegraphics[width=\columnwidth]{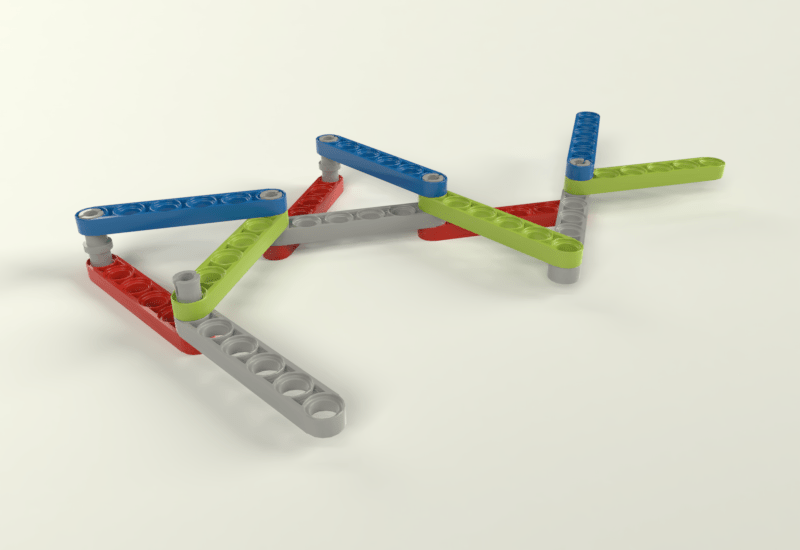}
  \caption{\label{lego_unit_cell}\strong{A possible unit cell to realize the Kagome lattice with LEGO.}
  In this LEGO realization, the unit cell has to be doubled in order for the vertices to satisfy the ice-like rule.
  The corresponding unit cell is composed of 12 \enquote{Liftarm 1 $\times$ 6 Thin} (32063) and 6 \enquote{Pin without Friction Ridges Lengthwise} (3673).
  To better show how the system is constructed, we attributed different colors to the beams at different heights.
  }
\end{figure}

Crystals on the two-dimensional Euclidean plane are classified according to their plane group (also called wallpaper group), the two-dimensional equivalent of space groups. To determine the plane group of a crystal, we use the library spglib \cite{spglib,Togo2018}, which automatically determines the space group of three-dimensional structures. To determine the plane group of a \emph{two}-dimensional structure, we embed the 2D crystal in the three-dimensional space, and periodize it in the $z$ direction orthogonal to the crystal plane (with an arbitrary period). This yields a three-dimensional crystal, the projection of which along the $z$ axis is the original plane structure. spglib can then be used to determine the space group of this 3D crystal.
Finally, we use the \enquote{symmetries of special projections} tabulated in the International Tables for Crystallography, Volume A \cite{ITA} to get the plane group of the two-dimensional projection along the original $z$ axis.
Table \ref{plane_groups_structures} provides a summary of the symmetry groups of the relevant structures. The initial Kagome lattice (with $\theta = 0$) has plane group \hm{p6mm}. The twisted Kagome $\theta \neq 0$ has plane group \hm{p31m}. We go from $(x,x,x,0)$ [\hm{p31m}] to $(x,x,x,z)$ [\hm{p3}] when $z \neq 0$, even if $z = z_c(x)$. 
In general, the deformed Kagome lattice has only plane group \hm{p1}. Note that despite having the simplest plane group \hm{p1}, the twisted Kagome lattice with three different masses still has a global two-fold degeneracy. 

Similar considerations apply to inequivalent bonds. To carry out the analysis, we insist that all bonds $B(M,M')$ connecting two classes of equivalent masses are equivalent (symbolically, $B(M,M_0) \sim B(M',M_0)$ as soon as $M \sim M'$). The situation of inequivalent bonds is then reduced to inequivalent masses. Note that this implies that we always have $B_1 \sim B_5$, $B_2 \sim B_6$, and $B_3 \sim B_4$ ; to describe a system with e.g. $k_1 \neq k_5$, a larger unit cell has to be chosen.

\section{LEGO realization of the mechanism}
\label{app_LEGO}

The LEGO realization of the Kagome lattice allows to demonstrate its collapse mechanism.
It is composed of LEGO Technic liftarms connected by pins, see figure~\ref{lego_unit_cell}.

Each pin can be attached to at most four liftarms, at different heights $h_1,\dots, h_4$ ; and in the Kagome lattice, it must be attached to exactly four liftarms.
As the liftarms are rigid, they cannot be bent, so the two pins connected by a liftarm should be attached at the same height.
Amusingly, this constraint is similar to the ice rule of the six-vertex model.
A practical consequence is that the unit cell of the Kagome lattice has to be enlarged in the LEGO realization.

\bibliography{bibliography}

\end{document}